\documentclass[10pt,conference]{IEEEtran}
\usepackage{ragged2e}
\usepackage{floatrow}
\usepackage{multirow}
\usepackage{amsthm}
\usepackage[utf8]{inputenc}
\usepackage[english]{babel}
\theoremstyle{remark}
\newtheorem{theorem}{Theorem}
\newtheorem{lemma}{Lemma}

\newtheorem{Proposition}{Proposition}

\usepackage{multicol}
\usepackage{epsfig}
\usepackage{epstopdf}
\usepackage{float}
\usepackage{perpage}
\MakeSorted{figure}
\MakeSorted{table}
\usepackage[normalem]{ulem}
\usepackage{array}
\usepackage{graphicx}
\usepackage{amsfonts}
\usepackage{amssymb}
\usepackage{soul}

\usepackage{enumitem}							
\usepackage[english]{babel}
\usepackage[utf8]{inputenc}
\usepackage[linesnumbered,ruled,vlined]{algorithm2e}
\usepackage{upgreek}
\usepackage{bm} 
\usepackage{hyperref}
\usepackage{float}
\floatstyle{plaintop}
\restylefloat{table}
\hypersetup{linktocpage} 
\hypersetup{
	colorlinks,
	citecolor=black,
	filecolor=black,
	linkcolor=black,
	urlcolor=black
}

\usepackage{pifont}
\usepackage{makecell}



\usepackage{cite}
\DeclareMathAlphabet\mathbfcal{OMS}{cmsy}{b}{n}
\ifCLASSINFOpdf
\else
\fi
\usepackage[cmex10]{amsmath}

\usepackage{array}
\usepackage{fixltx2e}

\usepackage[linesnumbered,ruled,vlined]{algorithm2e}

\usepackage[font=footnotesize,labelfont=bf, figurename=Fig.]{caption} 
\usepackage{subcaption}
\usepackage{fix-cm}
\usepackage{comment}

\usepackage{mathtools}
\usepackage{pifont}
\newcommand{\C}{\mathbb{C}}
\newcommand{\h}{\mathbf{h}}
\newcommand{\R}{\mathbf{R}}
\newcommand{\W}{\mathbf{W}}
\newcommand{\E}{\mathbb{E}}
\newcommand{\A}{\mathbf{A}}
\newcommand{\Amat}{\mathbf{A}}
\newcommand{\Bmat}{\mathbf{B}}
\newcommand{\Cmat}{\mathbf{C}}

\newcommand{\Hmat}{\mathbf{H}}
\newcommand{\Imat}{\mathbf{I}}
\newcommand{\Jmat}{\mathbf{J}}

\newcommand{\Pmat}{\mathbf{P}}

\newcommand{\Rmat}{\mathbf{R}}
\newcommand{\Smat}{\mathbf{S}}

\newcommand{\Vmat}{\mathbf{V}}
\newcommand{\Wmat}{\mathbf{W}}
\newcommand{\Xmat}{\mathbf{X}}
\newcommand{\Ymat}{\mathbf{Y}}
\newcommand{\Zmat}{\mathbf{Z}}
\newcommand{\av}{\mathbf{a}}
\newcommand{\bv}{\mathbf{b}}

\newcommand{\hv}{\mathbf{h}}

\newcommand{\nv}{\mathbf{n}}

\newcommand{\sv}{\mathbf{s}}

\newcommand{\vv}{\mathbf{v}}
\newcommand{\wv}{\mathbf{w}}
\newcommand{\xv}{\mathbf{x}}
\newcommand{\yv}{\mathbf{y}}

\newcommand{\phiv}{\boldsymbol{\phi}}
\newcommand{\gammav}{\boldsymbol{\gamma}}
\newcommand{\etav}{\boldsymbol{\eta}}

\newcommand{\Psimat}{\boldsymbol{\Psi}}

\newcommand{\Sigmamat}{\boldsymbol{\Sigma}}

\newcommand\eqa{\stackrel{\mathclap{\normalfont(a)}}{=}}

\IEEEoverridecommandlockouts\IEEEpubid{\makebox[\columnwidth]{ 978-1-6654-3540-6/\$31.00~22~\copyright~2022 IEEE \hfill} \hspace{\columnsep}\makebox[\columnwidth]{ }}
\begin{document}
	\bstctlcite{IEEEexample:BSTcontrol}
	\title{Hardware-Aware Pilot Decontamination Precoding for Multi-cell mMIMO Systems With Rician Fading}
	\author{\\[-25pt]
		\IEEEauthorblockN{Harshit Kesarwani, Dheeraj Naidu Amudala, Venkatesh Tentu and Rohit Budhiraja}
		\IEEEauthorblockA{Department of Electrical Engineering, Indian Institute of Technology Kanpur  
			\\\{harkes, dheeraja, tentu, rohitbr\}@iitk.ac.in\\[-25pt] 
			\thanks{This work is supported by Visvesvaraya PhD Scheme, MeitY, Govt. of India. MEITY-PHD-2721.}
		}
	}
	\maketitle
	\begin{abstract}
		We consider a hardware-impaired multi-cell Rician-faded massive multi-input multi-output (mMIMO) system with two-layer pilot decontamination precoding, also known as large-scale fading precoding (LSFP). Each BS is equipped with a flexible dynamic analog-to-digital converter (ADC)/digital-to-analog converter (DAC) architecture and the user equipments (UEs) have low-resolution ADCs. Further, both BS and UEs have hardware-impaired radio frequency chains. The dynamic ADC/DAC architecture allows us to vary the resolution of ADC/DAC connected to each BS antenna, and suitably choose~them to maximize the SE. We propose a distortion-aware minimum mean squared error (DA-MMSE) precoder and investigate its usage with two-layer LSFP and conventional single-layer precoding (SLP) for hardware-impaired mMIMO systems. We discuss the use cases of LSFP and SLP with DA-MMSE and distortion-unaware MMSE (DU-MMSE) precoders, which will provide critical insights to the system designer regarding their usage in practical systems. 
	\end{abstract}
	\vspace{-3pt}
	\section{Introduction}\vspace{-3pt}
	A massive multiple-input multiple-output (mMIMO) base station (BS) multiplexes a large number of user equipments (UEs) on the same spectral resource, which leads to a high spectral efficiency (SE)~\cite{Ozgocan_2019}. In a multi-cell mMIMO system, due to limited number of orthogonal pilots, UEs in different cells are forced to reuse the pilots. This leads to pilot contamination, which  reduces the SE~\cite{Ashikhmin_2018}.
	To mitigate the detrimental effects of pilot contamination, Ashikmin \textit{et al.} in~\cite{Ashikhmin_2018} proposed a two-layer large-scale fading precoding (LSFP) scheme. In the first layer of LSFP, a central network controller (CNC) generates a weighted combination of the messages of pilot-sharing UEs. The LSFP weights depend on the large-scale fading coefficients of the channels between the BS and a UE. The BSs later locally perform the second layer of precoding. The LSFP mitigates the effects of pilot contamination, outperforms the single-layer precoding (SLP) technique, which only performs local precoding.
	Demir \textit{et. al.}~in \cite{demir_large2020} considered the downlink of a spatially-correlated Rician-faded mMIMO system with phase-shifts, and designed optimal LSFP coefficients. 
	The authors in~\cite{Ozdogan_2019} and \cite{Van_2019} designed LSFD scheme (uplink counterpart of the LSFP scheme) for the uplink of  multi-cell and cell-free mMIMO systems by considering correlated Rayleigh and Rician fading,~respectively.
	
	For a cost-effective implementation of mMIMO systems, it is preferable to design them using low-cost radio-frequency (RF) chains and low-resolution {analog-to-digital converters (ADCs)/digital-to-analog converters(DACs)}. Such components, however, distort the transmit/receive signals~\cite{bjornson_da_2018,liu_2020,jacobsson_2017}. Bjornson \textit{et al.} in \cite{bjornson_da_2018} investigated the impact of correlated RF distortion in the uplink of a single-cell mMIMO system. The authors in \cite{liu_2020} considered low-resolution ADC and derived an asymptotic-SE expression for uncorrelated Rician-faded channels. Jacobsson \textit{et al.} in \cite{jacobsson_2017} showed that the SE achieved using infinite-resolution ADCs can be approached using few bits of resolution. The aforementioned single-cell mMIMO works considered either RF impairments~\cite{bjornson_da_2018} or low-resolution ADC/DACs~\cite{liu_2020,jacobsson_2017}.
	Xu \textit{et al.} in \cite{xu_2019} explored the joint impact of low-resolution ADCs and RF impairments.
	
	The authors in \cite{liang_2016,pirzadeh_2018,zhang_2017} proposed a mixed-resolution ADC architecture for single cell mMIMO systems, wherein the BS employs a combination of low- and high-resolution ADCs. Liang \textit{et al.} in \cite{liang_2016} showed that the mixed-resolution ADC architecture has a significantly higher SE than its low-resolution counterpart.
	{Pirzadeh \textit{et al.} in \cite{pirzadeh_2018} designed a channel estimator, while Zhang \textit{et al.} in \cite{zhang_2017} derived closed-form uplink SE expression for the mMIMO system with mixed-resolution architecture.}
	References \cite{bjornson_da_2018,liu_2020,jacobsson_2017,xu_2019,liang_2016,pirzadeh_2018,zhang_2017} considered single-cell uncorrelated uplink  mMIMO systems with Rayleigh fading, except \cite{zhang_2017}, which considered uncorrelated Rician fading. 
	
	
	
	

	The existing works in \cite{bjornson_da_2018, xu_2019,jacobsson_2017,liu_2020,liang_2016,pirzadeh_2018,zhang_2017} did not study the effect of low/mixed-resolution ADCs/DACs and low-cost RF chains on a multi-cell mMIMO system with LSFP/LSFD. The current work address these gaps for the downlink of a multi-cell mMIMO LSFP system with RF impairments and dynamic resolution architecture, which enables us to vary the resolution of each ADC/DAC from $1$ to $b$ bits. 
	Further, the combiners in \cite{xu_2019,jacobsson_2017,liu_2020,liang_2016,pirzadeh_2018,zhang_2017 } are unaware of hardware impairments. Their performance degrades in a harware-impaired mMIMO system. References~\cite{bjornson_da_2018,tugfe2019channel} investigated distortion-aware combiners for hardware-impaired mMIMO systems.
	The hardware-impaired mMIMO works in \cite{xu_2019,jacobsson_2017,liu_2020,liang_2016,pirzadeh_2018,zhang_2017,bjornson_da_2018,tugfe2019channel} considered Rayleigh fading channels, except \cite{zhang_2017,tugfe2019channel}, which considered Rician fading channels, but only with static phase-shifts in the line of sight (LoS) component. The static LoS phase-shift incorrectly models the  mobility and phase noise \cite{Ozdogan_2019}. To model them correctly, the authors in \cite{Ozdogan_2019,demir_large2020} considered a uniformly-distributed random phase-shift in the LoS component.  References \cite{Ozdogan_2019,demir_large2020}, however, assumed ideal hardware.
	The \textbf{main contributions} of the current work are next summarized as follows:
	\newline
	$\bullet$ We consider the downlink of a multi-cell spatially-correlated Rician-faded  mMIMO system with random LoS phase-shifts, and study the impact of dynamic-ADC/DAC architecture and cost-effective RF chains at the BS and UE. We extend the two-layer LSFP design from~\cite{Ashikhmin_2018,demir_large2020} to mitigate the interference due to  pilot contamination and hardware impairments. We also propose a distortion-aware (DA) MMSE precoder and demonstrate its gains over its distortion-unaware (DU) counterpart and maximal ratio (MR) precoding.
	\newline 
	$\bullet$ We maximize the non-convex sum-SE by optimizing the LSFP coefficients using minorization-maximization (MM) technique \cite{sun_2017}, which iteratively maximizes a non-concave objective using its concave surrogate lower bound function~\cite{sun_2017}.\newline
	$\bullet$ We provide tangible insights regarding the use of DA and DU precoders, and LSFP design in a hardware-impaired multi-cell mMIMO systems. We show that for low-to-moderate hardware impairments, the DA-MMSE with SLP, has almost~same SE as that of the DU-MMSE with LSFP. The proposed DA-MMSE precoder can thus help in avoiding the LSFP implementation, which requires the BS to exchange LSFP coded symbols over backhaul.  \textit{The current study, therefore, plays a key role in helping a system designer make an informed decision about the LSFP usage.} 
	\vspace{-3pt}
	\section{System Model}\vspace{-3pt}
	We consider the downlink of an $L$-cell mMIMO cellular network, with each cell consisting of an $M$-antenna BS and $K$ single-antenna UEs. The BSs are equipped with dynamic resolution ADC/DAC architecture and low-cost RF chains, while the UEs are equipped with low-resolution ADC/DAC and low-cost RF chains. The dynamic resolution ADC/DAC architecture enables each BS antenna to be connected to a different resolution ADC.~{This architecture is relevant for commercial 5G mMIMO deployment, as these systems, due to thermal constraints, are often designed as multiple 4/8 antenna subsystems which are then interfaced together~\cite{Jeon_Mag_2021}. The proposed dynamic architecture allows us to choose different resolution for each sub-system, and can achieve higher system SE.} 
	We next explain the channel model, and then the channel estimation and data transmission phases.
	\newline
	\underline{\textbf{Channel Model:}} 
	We denote the $k$th UE in the $l$th cell as $U_{lk}$. Its Rician-faded channel to the  $j$th BS i.e., $\mathbf{h}_{lk}^j \in \C^{M \times 1}$, is expressed as \cite{Ozdogan_2019,demir_large2020}:
	\begin{align}	
		\hv_{lk}^j={\bar{\hv}_{lk}^je^{j\phi_{lk}^j}}+\mathbf{R}_{lk}^{j^{\frac{1}{2}}}\mathbf{h}_{w_{lk}}^j. \label{h}
	\end{align}
	Here $\bar{\hv}_{lk}^j\triangleq\sqrt{\beta_{lk}^j \overline{K}_{lk}^j}\mathbf{{h}}_{m_{lk}}^j$ and $\R_{lk}^j\triangleq \beta_{lk}^j(1-\overline{K}_{lk}^{j}){\boldsymbol{\Sigma}}_{lk}^{j}$, with $\overline{K}_{lk}^{j} = K_{lk}^{j}/(K_{lk}^{j}+1)$. The scalar ${\phi_{lk}^j}$ in \eqref{h} is the random LoS phase-shift, which is uniformly distributed between $[-\pi,\pi]$~\cite{demir_large2020}. The vector $\mathbf{{h}}_{m_{lk}}^j$ in $\bar{\hv}_{lk}^{j}$ is the deterministic LoS component. The scalars $K_{lk}^{j}$ and $\beta_{lk}^{j}$ denote the Rician $K$-factor and the large-scale fading coefficient, respectively. 
	The matrix $\boldsymbol{\Sigma}_{lk}^{j} \in \C^{M \times M}$ in $\Rmat_{lk}^{j}$  represent the spatial correlation matrix of the channel $\mathbf{h}_{lk}^j$. The vector $\mathbf{h}_{w_{lk}}^j$, with $\mathcal{C}\mathcal{N}\left(0,1\right)$~elements, models small scale fading.
	\newline
	\underline{\textbf{Channel Estimation:}}
	The BSs estimate the uplink channel using the pilot signals transmitted by the UEs, and use the estimated uplink channels later to design its downlink precoders~\cite{demir_large2020}. We assume that all the UEs in each cell transmit orthogonal pilots, and UEs with same indices in different cells share the same pilot, which causes pilot contamination \cite{Ozgocan_2019}. Let $\phiv_k \in \mathbb{C}^{\tau_p\times 1}$ be the unit-power pilot signal of the $k$th UE in each cell, such that $\bm{\phi}_{k}^H\bm{\phi}_{k'}=\tau_p$, for $k=k'$ and $=0$ otherwise.
	The UE $U_{lk}$ scales its pilot signal $\bm{\phi}_{k}$ as $\sqrt{\tilde{p}_{lk}}\bm{\phi}_{k}$, with $\tilde{p}_{lk}$ being its transmit power. 
	This signal is fed to a low-resolution DAC, which distorts it. The distorted output signal of the UE $U_{lk}$, based on Bussgang model~\cite{Demir_2021}, is given as:
	\begin{align}
		\tilde{\mathbf{s}}_{DAC_{lk}}  = \alpha_{d_{lk}}\sqrt{\tilde{p}_{lk}}\boldsymbol{\phi}_{k}^T+\bar{\mathbf{n}}_{DAC_{lk}}^{T}.\vspace{-5pt}
	\end{align} 
	Here $\alpha_{d_{lk}} \!= (1 -\rho_{d_{lk}}) $ is the Bussgang gain, with $\rho_{d_{lk}}$~being the DAC distortion factor~\cite{Demir_2021}. 
	The vector $\bar{\mathbf{n}}_{DAC_{lk}}$ is~the quantization noise, whose elements have zero mean and variance  $\alpha_{d_{lk}}(1-\alpha_{d_{lk}})\tilde{p}_{lk}$. This quantized signal is then fed~to~the~low-cost hardware impaired RF chains. They further add a distortion noise $\bar{\boldsymbol{\eta}}_{tu_{lk}}^T$ which, according to the error vector magnitude (EVM) model, has zero mean and covariance $\kappa_{tu}^2 \E(\tilde{\sv}_{DAC_{lk}}\tilde{\sv}_{DAC_{lk}}^H)$~\cite{bjornson_da_2018}. Here $\kappa_{tu}$ characterizes the EVM of the UE transmit RF chain. The effective pilot signal transmitted by UE $U_{lk}$ is therefore $\big(\alpha_{d_{lk}}\sqrt{\tilde{p}_{lk}}\boldsymbol{\phi}_{k}^T\!+\!\bar{\mathbf{n}}_{DAC_{lk}}^{T}\!+\!\bar{\boldsymbol{\eta}}_{tu_{lk}}^T \big)$.
	The signal received at the antennas of the $j$th~BS is 
	\begin{align}
		\mathbf{Y}_{BS}^{j} \!=\! \sum_{l'=1}^{L}\sum_{k'=1}^{K}\hv_{l'k'}^{j}\!\left(\!\alpha_{d_{l'k'}}\sqrt{\tilde{p}_{l'k'}}\phiv_{k'}^T+\bar{\nv}_{DAC_{l'k'}}^{T}+\bar{\etav}_{tu_{l'k'}}^T\right).\notag
	\end{align}
	The signal received by the BS is fed to its low-cost RF chains  whose distorted output, based on the EVM model \cite{bjornson_da_2018}, is given as $\mathbf{Y}_{RF}^{j}=\mathbf{Y}_{BS}^{j}+\overline{\boldsymbol{\eta}}_{rb}^{j}+\overline{\mathbf{Z}}^{j}$. The matrices $\overline{\etav}_{rb}^{j}$ and $\overline{\Zmat}^{j}$ denote the receive distortion and AWGN at the $j$th BS. The columns of the matrix $\overline{\boldsymbol{\eta}}_{rb}^j$ are i.i.d. with zero mean and covariance $\kappa_{rb}^2\mathbf{W}^j$, where $\kappa_{rb}$ denotes the receive EVM of the $l$th BS and $\W^j=\text{diag}\big\{\E\big[\mathbf{y}_{BS}^{j}\mathbf{y}_{BS}^{j^H}|{\mathbf{h}_{l'k'}^{j}}\big]\big\}$, for $l'=1,\cdots,L$, $k'=1,\cdots,K$,
	\footnote{We assume that $l'=1,\cdots,L$, $k'=1,\cdots,K$ throughout the paper.} 
	with $\yv_{BS}^j$ being column of the matrix $\Ymat^{j}_{BS}$. The columns of $\overline{\Zmat}^{j}$ are i.i.d with pdf $  \mathcal{C}\mathcal{N}\left(\mathbf{0}_M,\sigma^2\mathbf{I}_M\right)$. The RF chain output $\mathbf{Y}_{RF}^{j}$ is then quantized using dynamic-resolution ADCs, which introduces quantization errors. Recall that this architecture enables to vary the resolution of each ADC from $1$-bit to its maximum $b$ bits. The quantized output, based on multi-dimensional Bussgang model~\cite{Demir_2021}, is given as 
	\begin{align}
		\mathbf{Y}_{\!\!ADC}^j&\!=\!\mathbb{Q}(\mathbf{Y}_{\!RF}^j)\!=\!\mathbf{A}^{j}_a\mathbf{Y}_{\!RF}^{j}+\overline{\mathbf{N}}_{q}^{j}. \label{ADC}
	\end{align}
	The matrix $\Amat^j_a=\text{diag}(\alpha_{1}^j,\cdots\!,\alpha_{M}^j)$, with $\alpha_{i}^j$ denoting the Bussgang gain for the $i$th antenna of $j$th BS. The vector $\mathbf{N}_{q}^j$ is the additive ADC quantization noise whose i.i.d. columns have zero mean and covariance $\mathbf{B}^{j}_a\mathbf{S}^j$. Here $\mathbf{B}^{j}_a=\mathbf{A}^j_a(\mathbf{I}_M-\mathbf{A}^j_a)$, $\mathbf{S}^j = \text{diag}\big(\mathbb{E}\big[ \mathbf{y}_{\!RF}^j\mathbf{y}_{\!RF}^{j^H}\big|{\mathbf{h}_{l'k'}^{l}} \big] \big)$ and $\yv^{j}_{RF}$ is a column of the matrix $\Ymat^{j}_{RF}$. 
	The pilot signal received at the $j$th BS is given~as 
	\begin{align}
		\mathbf{Y}^{j}&\!=\!\sum_{l'=1}^{L}\sum_{k'=1}^{K}\mathbf{A}^{j}_a\hv_{l'k'}^{j}\! \left(\!\alpha_{d_{l'k'}}\sqrt{\tilde{p}_{l'k'}}\boldsymbol{\phi}_{k'}^T+\bar{\mathbf{n}}_{DAC_{l'k'}}^{T}+\bar{\boldsymbol{\eta}}_{tu_{l'k'}}^T \!\right)\! \nonumber\\[-3pt]
		&\hspace{20pt}+\mathbf{A}^{j}_a\bar{\boldsymbol{\eta}}_{rb}^j+\mathbf{A}^j_a\overline{\mathbf{Z}}^j+\overline{\mathbf{N}}_{q}^j \label{uplink signal rx}.
	\end{align}  
	\begin{figure*}[t]\vspace{2pt}
		\setcounter{equation}{10}
		{\small\begin{align} \label{LSFP_brief}
				{y}_{ADC_{lk}} 
				&= \underbrace{\alpha_{a_{lk}}\sum_{r=1}^{L}\mathbf{h}_{lk}^{r^H}\mathbf{A}_d^r\mathbf{w}^{r}_{rk}\gamma_{lk}^{r*}s_{lk}}_{\text{desired signal}} + \underbrace{\alpha_{a_{lk}}\sum_{r \neq l }^{L}\sum_{n=1}^{L}\mathbf{h}_{lk}^{r^H}\mathbf{A}_d^n\mathbf{w}^{n}_{nk}\gamma_{rk}^{n*}s_{rk}}_{\text{pilot contamination}} +\underbrace{\alpha_{a_{lk}}\sum_{r=1}^{L}\sum_{ k^{'} \neq k }^{K}\sum_{n=1}^{L}\mathbf{h}_{lk}^{r^H}\mathbf{A}_d^n(\mathbf{w}^{n}_{nk^{'}})\gamma_{rk^{'}}^{n*}s_{rk^{'}}}_{\text{non-pilot co-channel interference}} \nonumber\\[-3pt] &\hspace{30pt}+ \underbrace{\alpha_{a_{lk}}\sum_{r=1}^{L}\mathbf{h}_{lk}^{r^H}\boldsymbol{\eta}_{tb}^{r}}_{\text{BS transmit RF impairments}} + \underbrace{\alpha_{a_{lk}}\sum_{r=1}^{L}\mathbf{h}_{lk}^{r^H}\mathbf{n}^r_{DAC}}_{\text{DAC impairments}} +\underbrace{\alpha_{a_{lk}}\eta_{ru_{lk}}}_{\text{UE receive RF impairments}} + \underbrace{\alpha_{a_{lk}}n_{lk}}_{\text{noise}} + \underbrace{n_{ADC_{lk}}}_{\text{ADC impairments}}.
		\end{align}}
		\hrule
		\vspace{-14pt}
		\setcounter{equation}{4}
	\end{figure*}
	The $j$th BS estimates $\mathbf{h}_{lk}^{j}$, which is the channel of UE $U_{lk}$, by correlating its received signal~\eqref{uplink signal rx} with the pilot signal $\bm{\phi}_{k}$ as follows: $\mathbf{y}_{jk} = \mathbf{Y}^{j}\bm{\phi}_{k}^{*} = \sum\limits_{l'= 1}^L \alpha_{d_{l'k}}\sqrt{\tilde{p}_{l'k}}\tau_p\mathbf{A}^j_a\mathbf{h}_{l'k}^j + \sum\limits_{l'=1}^L\!\sum\limits_{k'=1}^K \mathbf{A}^{j}_a\mathbf{h}_{l'k'}^j\left(\!\bar{\mathbf{n}}_{DAC_{l'k'}}^T\!\!+\! \bar{\bm{\eta}}_{tu_{l'k'}}^T\!\right)\bm{\phi_{k}}^*
	\!+\mathbf{A}^{j}_a\bar{\boldsymbol{\eta}}_{rb}^j\boldsymbol{\phi}_{k}^*+\overline{\mathbf{N}}_{q}^{j}\boldsymbol{\phi}_{k}^* +\mathbf{A}^j_a\overline{\mathbf{Z}}^j\boldsymbol{\phi}_{k}^*$.
	In a practical system, the BS is unaware of the phases ~\cite{Ozdogan_2019,demir_large2020}.
	We next derive a phase-unaware linear MMSE (LMMSE) channel estimate of $\hv_{lk}^{j}$ in the following theorem. The proof is provided in~Appendix~\ref{app:chan_est}.
	\begin{theorem}
		The phase-unaware LMMSE estimate of the uplink  correlated Rician-faded channel $\hat{\mathbf{h}}_{lk}^{j}$ with dynamic-resolution ADC architecture and RF impairments, is given as\vspace{-2pt}
		\begin{align}
			\hat{\mathbf{h}}_{lk}^{j}=\alpha_{d_{lk}}\sqrt{\tilde{p}_{lk}}\mathbf{A}^{j}_a\overline{\mathbf{R}}_{lk}^j\boldsymbol{\Psi}_{jk}^{-1}\mathbf{y}_{jk}, \text{ where }  \label{estimate} 
		\end{align}
		$\overline{\mathbf{R}}_{lk}^j \!=\! \Rmat_{lk}^j + \overline{\hv}_{lk}^j(\overline{\hv}_{lk}^j)^{H}\!$ and $\Psimat_{jk}\!=\!\!\sum\limits_{l'=1}^L\alpha_{d_{lk}}^2\tau_p\tilde{p}_{lk}\Amat^{j}_a\overline{\Rmat}_{l'k}^j\Amat^{j^H}_a\!+\!\sum\limits_{l'=1}^L\sum\limits_{k'=1}^K \alpha_{d_{l'k'}}(1\!-\!\alpha_{d_{l'k'}}+\kappa_{tu}^2)\tilde{p}_{l'k'}\Amat^j_a\overline{\Rmat}_{l'k'}^j\Amat^{j^H}_a +\sigma^2\Amat^{j}_a\Amat^{j^H}_a$\\$ +\kappa_{rb}^2\A^j_a{\Jmat}^j\Amat^{j^H}_a+\mathbf{B}^j_a\left(\left(1+\kappa_{rb}^2\right){\mathbf{J}}^j+\sigma^2\mathbf{I}_M)\right)$. Due to LoS phase-shift, the channel estimate and the estimation error are uncorrelated, but not independent.
	\end{theorem}
	\hspace{-12pt}\underline{\textbf{Downlink Data Transmission:}} We adopt a two-layer precoding strategy, wherein the first layer the CNC performs LSFP based on the long-term statistics and in the second layer, the BSs perform local precoding  using the locally estimated channels. We now explain these two steps.
	\subsubsection*{\underline{Layer I -- LSFP precoding at the CNC}} Recall that the CNC knows the symbols of all the UEs. It performs LSFP by linearly combining the symbols transmitted by the pilot sharing UEs. If $s_{rk}$ is the zero-mean and unit-variance data symbol of the $k$th UE in the $r$th cell, then the linearly combined LSFP signal generated by the CNC for the $k$th UE of the $l$th cell is \cite{demir_large2020} $v_{lk} = \sum_{r=1}^L\gamma_{rk}^{l^*}s_{rk}$.
	Here $\gamma_{rk}^{l}$ is the complex LSFP weight which satisfy the transmit power constraint of the BS i.e., $\sum_{k = 1}^{K}\sum_{l=1}^{L}|\gamma_{lk}^r|^2 \leq \rho_d$, for $r = 1,\cdots,L$. They are designed later to maximize the SE.
	\subsubsection*{\underline{Layer II -- Local precoding at the BS}} The CNC transmits the LSFP-encoded symbols to the BSs, which locally perform second layer of precoding using the estimated channels. The local precoder is designed to mitigate the multi-UE interference from the UEs in the same/neighboring cells. The locally-precoded signal transmitted by the $l$th BS is given as
	\begin{align}
		&\mathbf{x}_l = \sum_{k=1}^K\mathbf{w}^{l}_{lk}v_{lk} = \sum_{k=1}^{K}\sum_{r = 1}^{L} \wv_{lk}^{l} \gamma_{rk}^{l^\ast} s_{rk}.
	\end{align}
	Here $\mathbf{w}^{l}_{lk} \in \C^{M \times 1}$ is the local precoder, whose design is discussed later in the sequel. 
	\subsubsection*{\underline{Hardware-impaired BS transmit signal}}
	The BS feeds the precoded transmit signal $\xv_l$ to the dynamic-architecture DACs, which add quantization noise. The quantized DAC output, based on the Bussgang model, is expressed as~\cite{Demir_2021}:
	\begin{align}
		\mathbf{x}^l_{DAC} = \mathbb{Q}(\mathbf{x}_l) = \mathbf{A}_d^l\mathbf{x}_l + \mathbf{n}^l_{DAC}.
	\end{align}
	The matrix $\mathbf{A}_d^l=\text{diag}\{\alpha_{d_1}^l,\cdots\!,\alpha_{d_M}^l\}$, with $\alpha_{d_i}^l$ being the Bussgang gain at the $i$th antenna of the $l$th BS. The vector $\mathbf{n}_{DAC}^l$ is the additive DAC quantization noise with zero mean and conditional covariance matrix $\mathbf{B}_d^{l}\mathbf{D}^l$ {\cite{Demir_2021}}. Here $\mathbf{B}_d^{l}=\mathbf{A}_d^l(\mathbf{I}_M-\mathbf{A}_d^l)$ and $\mathbf{D}^l = \text{diag}\left(\mathbb{E}\left[ \mathbf{x}_l \mathbf{x}_l^{H} |{\mathbf{h}_{l^{'}k^{'}}^{l}}\right ] \right)$.  The BS feeds the DAC output signal to the low-cost hardware-impaired transmit RF chains which further distorts it by adding a zero mean noise $\etav_{tb}^{l}$, with covariance $\E[\etav_{tb}^{l}\etav_{tb}^{l^H}] = \kappa_{tb}^2 \E[\xv_{DAC}^{l}\xv_{DAC}^{l^H}]$~\cite{bjornson_da_2018}. The effective signal transmitted by the $l$th BS, based on the EVM model~\cite{bjornson_da_2018} is, therefore 
	\begin{align}
		\mathbf{x}^l_{RF} = \mathbf{A}_d^l\mathbf{x}_l + \mathbf{n}^l_{DAC} + \boldsymbol{\eta}_{tb}^{l}.
	\end{align}
	\subsubsection*{\underline{Hardware impaired UE receive signal}} Each UE receives the signal transmitted by all the BSs. Recall that the UEs have a hardware-impaired RF chain and a low-resolution ADC. The RF chain output of the UE $U_{lk}$ is given as~\cite{bjornson_da_2018}:
	\begin{align}
		{y}_{RF_{lk}} = \sum_{r=1}^{L}\mathbf{h}_{lk}^{r^H}\mathbf{x}^r_{RF} + \eta_{ru_{lk}} + n_{lk}.
	\end{align}
	The AWGN $n_{lk}$ has pdf $\mathcal{CN}\!\left(0,\sigma^2\right)$, while the  
	hardware distortion $\eta_{ru_{lk}}\!$ has zero mean and~variance $\kappa_{ru}^2\delta_{lk}$, with  $\delta_{lk}=\text{diag}\big\{\mathbb{E}\big[\big(\sum\limits_{r=1}^{L}\mathbf{h}_{lk}^{r^H}\mathbf{x}^r_{RF}\big)\big(\sum\limits_{r=1}^{L}\mathbf{h}_{lk}^{r^H}\mathbf{x}^r_{RF}\big)^H\big|\hv_{lk}^{r}\big]\big\}$. The distorted RF chain~output is fed to the low-resolution ADC, whose quantized output,  based on Bussgang model {\cite{Demir_2021}}, is 
	\begin{align}
		{y}_{ADC_{lk}} =\mathbb{Q}({y}_{RF_{lk}}) = \alpha_{a_{lk}}{y}_{RF_{lk}} + n_{ADC_{lk}}.
	\end{align}
	The scalar $\alpha_{a_{lk}}$ is the Bussgang gain. The additive ADC quantization noise $n_{ADC_{lk}}$ has zero mean and variance  $\alpha_{a_{lk}}\left(1-\alpha_{a_{lk}}\right)\epsilon_{lk}$ such that $\epsilon_{lk}=\text{diag}\left\{\mathbb{E}\left[ \left|{y}_{RF_{lk}}\right|^2 \big|\mathbf{h}_{lk}^{r}\right]\right\}$.  We now decompose the signal ${y}_{ADC_{lk}}$ to show various interference and noise terms as in \eqref{LSFP_brief} at the top of this~page. 
	
	In \eqref{LSFP_brief}, the pilot contamination interference occurs due to sharing of pilots by UEs in different cells, while non-pilot co-channel interference is caused by the non pilot sharing UEs in all the cells. The remaining interference terms model the distortions of various UE and BS hardware impairments.  
	\newline 
	\underline{\textbf{Local-precoder designs:}} We now design various local precoders to be used at the BS. Note that the receive signal of the UE $U_{lk}$ in \eqref{LSFP_brief}, depends on the precoders of all the UEs in the network i.e., $\wv_{l'k'}^{l'}$ $\forall l',k'$. This makes downlink precoder design a non-trivial task~\cite{Ozdogan_2019,Ozgocan_2019,demir_large2020}. Existing mMIMO literature \cite{demir_large2020,Ozgocan_2019,Ozdogan_2019} commonly designs downlink precoders from the uplink combiners by leveraging the uplink-downlink duality principle~\cite{Ozgocan_2019,demir_large2020}. Two widely popular downlink precoders in the multi-cell mMIMO literature are~\cite{tugfe2019channel,Ozgocan_2019}:\vspace{-2pt}
	\begin{itemize}[leftmargin = *]
		\item \textit{MR precoder:} $\mathbf{w}_{lk}^l= \omega_{lk} \hat{\hv}_{lk}^{l}$, with $\omega_{lk}^2 = 1/\E\{\|\hat{\mathbf{h}}_{lk}^{l}\|^2\}$.
		\item \textit{DU-MMSE precoder:} $\wv_{lk}^l = \tilde{\omega}_{lk} \vv_{lk}$,  where $\tilde{\omega}_{lk}^2 = 1/\E\{\|\vv_{lk}\|^2\}$ and $\vv_{lk}$ being the $k$th column of matrix $\Vmat_{l}^{\text{du}} \!=\! \bigg(\!\widehat{\Hmat}_l^{H}\mathbf{P}_{l}\widehat{\mathbf{H}}_l \!+\! \sum\limits_{i=1}^{K}p_{li}\mathbf{C}_{li}^{j} + \sum\limits_{l'\neq l}^{L}\sum\limits_{i=1}^{K}p_{l'i}\mathbf{R}_{l'i}^{j} \!+\! \sigma^2\Imat\bigg)^{-1}\!\!\!\widehat{\mathbf{H}}_l\mathbf{P}_l$, with $\Pmat_l = \text{diag}\left(\tilde{p}_{l1},\cdots,\tilde{p}_{lK}\right)$ and $\widehat{\Hmat}_l = [\hat{\hv}_{l1}^{l},\cdots,\hat{\hv}_{lK}^{l}]$.
	\end{itemize}
	The MR and DU-MMSE precoder  precoders, however, crucially ignore the distortion caused by non-ideal BS and UE hardware. In a hardware-impaired mMIMO system, to realize tangible SE gains, it is crucial to mitigate the effect of hardware impairments along with different interferences~\cite{bjornson_da_2018}. We now propose a novel distortion-aware precoder, which exploits  the statistical knowledge about hardware impairments. Its proof is provided in~Appendix~\ref{app:DAMMSE}. 
	\begin{Proposition}
		For a hardware-impaired multi-cell mMIMO system, a distortion-aware MMSE precoder that mitigates the detrimental effect of imperfect hardware is given as \setcounter{equation}{11} \begin{align}\label{da mmse predoder exp}
			\mathbf{V}^{\text{da}}_{l} \!&=\! [\vv_{l1}^{\text{da}},\cdots,\vv_{lK}^{\text{da}}] = 
			{\widehat{\Vmat}_{l}}^{-1}\widehat{\mathbf{H}}_l\mathbf{P}_l  \text{ and } 
			\wv_{lk}^l = \widehat{\omega}_{lk} \mathbf{v}_{lk}^{\text{da}}.
		\end{align}
		Here $\widehat{\omega}_{lk}^2 \!=\! 1/\mathbb{E}\{\|\mathbf{v}_{lk}^{\text{da}}\|^2\}$ and ${\widehat{\Vmat}}^l$ is defined in~Appendix~\ref{app:DAMMSE}.
	\end{Proposition}
	\begin{figure*}[t]\vspace{2pt}
		\setcounter{equation}{14}
		{\small\begin{align}
				\overline{\text{SINR}}_{lk} &=\frac{\vphantom{\sum\limits^K}\alpha_{a_{lk}}^2\left|\boldsymbol{\gamma}_{lk}^{H}\bv_{lk}\right|^2}
				{\begin{Bmatrix}\alpha_{a_{lk}}^2\sum\limits_{\substack { r=1 }}^{L}\mathbf{{\boldsymbol{\gamma}}}_{rk}^H\mathbf{C}_{lkk}\mathbf{{\boldsymbol{\gamma}}}_{rk} \! - \alpha_{a_{lk}}^2\!\left|\boldsymbol{\gamma}_{lk}^{H}\bv_{lk}\right|^2 \! + \!\alpha_{a_{lk}}^2\sum\limits_{r=1}^{L}\sum\limits_{\substack { k^{'} \neq k }}^{K}\mathbf{{\boldsymbol{\gamma}}}_{rk'}^H\mathbf{C}_{lkk^{'}}\mathbf{{\boldsymbol{\gamma}}}_{rk^{'}} + \alpha_{a_{lk}} \sigma^2 \\[-3pt]
						+ \sum\limits_{n=1}^{L}\sum\limits_{r=1}^{L}\sum\limits_{k'=1}^{K}\!
						\alpha_{a_{lk}}\!\!\left[\sum\limits_{r'=1}^{L} (1+\kappa_{ru}^2- \alpha_{a_{lk}})\gamma_{nk'}^{r^*}\gamma_{nk'}^{r'^*}c_{lkk'}^{rr'}\!+\!(1+\kappa_{ru}^2)|\gamma_{nk^{'}}^r|^{2}\big[ e_{lkk'}^r+\kappa_{tb}^2 f_{lkk'}^r\big]\right]
				\end{Bmatrix}}\triangleq \frac{N_{lk}(\overline{\gammav})}{D_{lk}(\overline{\gammav})}.\label{SINR_frac}
		\end{align}} 
		\hrule
		\vspace{-13pt}
		\setcounter{equation}{12}
	\end{figure*} 
	\vspace{-5pt}
	\section{Achievable SE and LSFP optimization}\vspace{-2pt}
	We now exploit the “hardening-bound”  technique to derive a SE lower bound. We re-express the signal in \eqref{LSFP_brief} as~\cite{demir_large2020}:
	\begin{align} \label{s_lower}
		y_{ADC_{lk}} \!&= \underbrace{\alpha_{a_{lk}}\!\sum_{r=1}^{L}\E\{\mathbf{h}_{lk}^{r^{H}}\mathbf{A}_d^r\mathbf{w}^{r}_{rk}\}\gamma_{lk}^{r*}s_{lk}}_{\text{desired signal}} + \underbrace{\tilde{z}_{lk}}_{\text{effective noise}}\!.
	\end{align}
	The effective noise term $\tilde{z}_{lk}$ includes all the interference terms  in \eqref{LSFP_brief} plus beamforming uncertainty given as  $\alpha_{a_{lk}}\sum\limits_{r=1}^{L}{\gamma}_{lk}^{r^*}\left(\mathbf{h}_{lk}^{r^{H}}\mathbf{A}_{d}^r\mathbf{w}_{rk}^r-\E\{\mathbf{h}_{lk}^{r^{H}}\mathbf{A}_{d}^r\mathbf{w}_{rk}^r\}\right)s_{lk}$.
	The desired signal and the effective noise $\tilde{z}_{lk}$ can easily shown to be uncorrelated. The effective noise $\tilde{z}_{lk}$  is the sum of many terms, which enables us to treat it as worst-case Gaussian noise \cite{Ozgocan_2019}. This helps in proposing the following tight SE lower bound:
	\begin{align}
		\overline{\text{SE}}_{\text{sum}} &= \frac{\tau_c-\tau_p}{\tau_c} \sum_{l=1}^{L}\sum_{k=1}^{K} \log_2\left(1 + \overline{\text{SINR}}_{lk}\right),  \label{SE_bar}
	\end{align}
	where $\overline{\text{SINR}}_{lk}$ is given in \eqref{SINR_frac} at the top of next~page. In \eqref{SINR_frac}, $\gammav_{lk} = [\gammav_{lk}^{1},\cdots,\gammav_{lk}^{L}]^T\!\in\!\C^{L\times 1}$ and $\overline{\gammav} \!=\! [\gammav_{11}^T,\!\cdots\!,\gammav_{1K}^T,\cdots,\gammav_{L1}^T,\cdots,\gammav_{LK}^T]^T$. The terms $\bv_{lk} = [b_{lk}^{1},\cdots,b_{lk}^{L}]^T$, $\mathbf{C}_{lkk'} = \left[  c_{lkk'}^{11},\cdots,c_{lkk'}^{1L};c_{lkk'}^{21},\right.$ $\left.\cdots,c_{lkk'}^{2L};\cdots; c_{lkk'}^{L1},\cdots,c_{lkk'}^{LL}\right]$, $e_{lkk'}^{r}$, $f_{lkk'}^{r}$ are defined as
	\begin{align}
		&b_{lk}^{j} \!=\! \E\{\mathbf{h}_{lk}^{r^{H}}\mathbf{A}_{d}^r\mathbf{w}_{rk}^{r}\},\;c_{lkk'}^{mn} \!=\! \E\left\{\!\hv_{lk}^{m^{H}}\!\Amat_{d}^m\wv_{mk'}^m\wv_{nk}^{n^{H}}\Amat_{d}^n\hv_{lk'}^{n}\!\right\},\nonumber
	\end{align}
	$e_{lkk'}^{r} = \breve{e}_{lkk'}^{r}(\Bmat_{d}^{r})$ and $f_{lkk'}^{r} = \breve{e}_{lkk'}^{r}(\Amat_{d}^{r})$, with $ \breve{e}_{lkk'}^{r}(\Xmat) = \E\left\{\hv_{lk}^{r}\Xmat\text{diag}\big(\wv_{rk'}^{r}\wv_{rk'}^{r^H}\big) \hv_{lk}^{r}\right\}$.\\
			In denominator of \eqref{SINR_frac}, the first two terms are beamforming uncertainty plus pilot contamination power, the third term is the non-pilot co-channel interference power, the last two terms represent the powers of BS and UE impairments. \newline	
			\underline{\textbf{LSFP optimization:}}
			We now design the LSFP coefficients to maximize the SE of a hardware-impaired multi-cell mMIMO system. The SE maximization problem is given as 
			\begin{align}\label{optimization_mm}\setcounter{equation}{15}
				\mathbf{P1}: &\underset{\overline{\gammav}}{\text{ Maximize }}
				\sum\limits_{l=1}^{L}\sum\limits_{k=1}^{K}\log_2\left( 1 + \frac{N_{lk}(\overline{\gammav})}{D_{lk}(\overline{\gammav})}\right) \triangleq f_{\text{SE}}(\overline{\gammav}), \nonumber\\
				&\text{ subject to }
				\sum\limits_{k=1}^{K}\sum\limits_{l=1}^{L}|\mathbf{\gamma}_{lk}^{r}|^2 \leq\rho_d, \forall r.
			\end{align}\vspace{-1pt}
			The constraint in \eqref{optimization_mm} is the transmit power budget at each BS. The SE metric consists of fractions $N_{lk}(\overline{\gammav})/D_{lk}(\overline{\gammav})$, with $N_{lk}(\overline{\gammav})$ and $D_{lk}(\overline{\gammav})$ containing product of optimization variables, which makes it non-convex.
			We solve it using MM algorithm \cite{sun_2017}, which considers the following problem:
			$\underset{\xv \in \mathcal{X}}{\text{maximize }} f(\xv) $.
			The MM algorithm has  minorization and  maximization steps. In the former step, it constructs a surrogate function $g(\xv|\hat{\xv}_t)$ to lower bound the  objective  $f(\xv)$ at a feasible point $\xv_t$. In the latter step, it finds a feasible point by maximizing the surrogate as    $\hat{\xv}_{t+1} = \underset{\xv}{\text{argmax}} g(\xv|\hat{\xv}_t)$.
			
			The MM algorithm, thus, iteratively generates a sequence of feasible points $(\hat{\xv}_{{t}}|t\in\!\mathbb{N})$ which converge to a stationary solution of the original problem. A surrogate function is valid if it additionally satisfies the following two
			conditions~\cite{sun_2017}:
			\begin{align}\label{opt_conditions}
				\!\!\!\!\!g(\hat{\mathbf{x}}_{t}|\hat{\mathbf{x}}_{t}) = f(\hat{\mathbf{x}}_{t}) \text{ and } \nabla_{\mathbf{x}}g(\mathbf{x}|\mathbf{\hat{x}}_{t})|_{\mathbf{x} = \mathbf{\hat{x}}_{t}} = \nabla_{\mathbf{x}}f(\mathbf{x})_{\mathbf{x}=\mathbf{\hat{x}}_{t}}. 
			\end{align}
			\underline{\textbf{Surrogate function construction:}}
			We will construct a valid surrogate function for the non-concave objective of $\mathbf{P1}$ i.e., $f_{\text{SE}}(\overline{\gammav})$. Recall that SE is sum of multiple ratios, further:\vspace{-7pt}
			\begin{itemize}[leftmargin = *]
				\item the numerator in each ratio $\sqrt{N_{lk}(\overline{\gammav})} \!=\! \sqrt{\gammav_{lk}^H\bv_{lk}\bv_{lk}^{H}\bm{\gamma}_{lk}}$ is a composition of square root and a convex function, which is non-convex according to the composition~rules~\cite{Boyd_2014}.
				\item the denominator term $D_{lk}(\overline{\gammav})$ consists of product of optimization variables $\gamma_{nk'}^{r}\gamma_{nk}^{r'}$ for $(r',k')\neq (r,k)$ (see fifth term in the denominator of \eqref{SINR_frac}).
			\end{itemize} 
			We now state two lemmas to handle the non-convexities, and eventually lower-bound the SE. {Their proofs are given in Appendix~\ref{app:MM_cond}.}
			\begin{lemma}\label{surrogate_fun}
				Let  $A_{lk}(\mathbf{x}):\mathbb{R}^{n} \rightarrow \mathbb{R}_{+}$ and $B_{lk}(\mathbf{x}):\mathbb{R}^{n}\rightarrow \mathbb{R}_{++} (\forall l=1,\cdots,L \text{ and } k=1,...K)$, be a positive function and a non-negative function, respectively. At any feasible point $\mathbf{x} = \mathbf{x}_t$, a surrogate function that lower bounds $\frac{A_{lk}(\xv)}{B_{lk}(\xv)}$ is:
				\begin{align}
					\frac{A_{lk}(\mathbf{x})^2}{B_{lk}(\mathbf{x})}\geq 2y_{lk}A_{lk}(\mathbf{x})  - y_{lk}^2B_{lk}(\mathbf{x}) = h_{lk}(\xv,\xv_{t}).
				\end{align}
				Here $y_{lk}$ is a function of $\xv_t$ and is calculated as $y_{lk} \! =\!  \frac{A_{lk}(\mathbf{x}_t)}{B_{lk}(\mathbf{x}_t)}$. 
			\end{lemma}
			\begin{lemma}\label{lemma_linear_surrogates}
				Given a non-concave function  $f_1(\bm{\gamma}_{lk}) = \sqrt{\bm{\gamma}_{lk}^H\bv_{lk}\bv_{lk}^{H}\bm{\gamma}_{lk}}$ and $f_2(\gamma_{nk}^{r'},\gamma_{nk}^{r}) = \gamma_{nk}^{r'}\gamma_{nk}^{r}$, we can construct a linear surrogate function $g_1(\bm{\gamma}_{lk})$ and $g_2(\gamma_{nk}^{r'},\gamma_{nk}^{r})$ at $\gammav_{lk}^{(t)}$ and $(\gamma_{lk}^{r^{(t)}},\gamma_{lk}^{r'^{(t)}})$ respectively, by using Taylor's first order expansion as \label{taylor_surrogate}
				\begin{align}
					&g_1(\boldsymbol{\gamma}_{lk}) = \sqrt{\boldsymbol{\gamma}_{lk}^{(t)^{H}}\!\bv_{lk}\bv_{lk}^H{\gammav}_{lk}^{(t)}} \!+\!\frac{\gammav_{lk}^{(t)^{H}}\!\bv_{lk}\bv_{lk}^H(\gammav_{lk} - \gammav_{lk}^{(t)})}{\sqrt{\gammav_{lk}^{(t)^{H}}\bv_{lk}\bv_{lk}^H\gammav_{lk}^{(t)}}} \text{ and } \nonumber\\
					&g_2(\gamma_{lk}^{r}\gamma_{lk}^{r'}) =  \gamma_{lk}^{r^{(t)}}\gamma_{lk}^{r'}+\gamma_{lk}^{r'^{(t)}}\gamma_{lk}^{r}-\gamma_{lk}^{r^{(t)}}\gamma_{lk}^{r'^{(t)}}. \nonumber 
				\end{align}
			\end{lemma}
			The functions $h_{lk}(\xv)$, $g_1(\gammav_{lk})$ and $g_2(\gamma_{lk}^{r}\gamma_{lk}^{r'})$, can be easily shown to satisfy the aforementioned necessary conditions in \eqref{opt_conditions}, and are therefore valid surrogate functions. We now use Lemma \ref{surrogate_fun} to decouple the multiple fractions and Lemma~\ref{lemma_linear_surrogates} for the non-convex terms in the numerator and denominators of the sum SE metric. The equivalent lower bound, at a  feasible point $\overline{\gammav} = \overline{\gammav}^{(t)}$, is given as $f_{\text{SE}}(\overline{\gammav})\geq \breve{f}_{\text{SE}}(\overline{\gammav};\overline{\gammav}^{(t)})$, where
			\begin{align}\label{mm_opt_cvx}		\!\!\!\!\!\!\!\breve{f}_{\text{SE}}(\overline{\gammav};\overline{\gammav}^{(t)}) \!=\! \sum\limits_{l=1}^{L}\sum\limits_{k=1}^{K}\!\log_2 \!\Big(\!1\!+\!2z_{lk}\!\sqrt{\!\widetilde{N}_{lk}(\overline{\gammav})} \!- \!z_{lk}^2\widetilde{D}_{lk}(\overline{\gammav})\!\Big).
			\end{align}\vspace{-2pt}
			The scalar $z_{lk}$, using Lemma~\ref{surrogate_fun}, is given as $z_{lk}= \sqrt{\!N_{lk}\!(\overline{\gammav}^{(t)})}/D_{lk}(\overline{\gammav}^{(t)})$.	
			Here $\widetilde{N}_{lk}(\overline{\gammav})$ (resp. $\widetilde{D}_{lk}(\overline{\gammav})$) is linear surrogate function over $N_{lk}(\overline{\gammav})$ (resp. $D_{lk}(\overline{\gammav})$) obtained using Lemma~\ref{taylor_surrogate}. The problem $\mathbf{P1}$, by using \eqref{mm_opt_cvx}, is recast as\vspace{2pt}
			\begin{align}\label{P2}
				\!\!\!\!\!\!\mathbf{P2}: &\underset{{\overline{\gammav}}}{\text{ Maximize }}
				\breve{f}_{\text{SE}}(\overline{\gammav};\overline{\gammav}^{(t)}), \!\text{ s.t. }\!
				\sum\limits_{k=1}^{K}\sum\limits_{l=1}^{L}|\mathbf{\gamma}_{lk}^{r}|^2 \leq\rho_d, \forall r.
			\end{align}
			Problem $\mathbf{P2}$ is now concave in $\overline{\gammav}$, which can be solved using MM algorithm given in~Algorithm~\ref{MMalgo}. 
			\begin{algorithm}[h!]
				\footnotesize	
				\textbf{Input:} Choose a suitable tolerance $\epsilon > 0$ and initialize $\overline{\gammav}^{(0)}$ such that it satisfies the per BS power constraint in \eqref{P2}. \\
				\textbf{Output:} $\overline{\gammav}^{\ast}$\\
				\For {$t \leftarrow 1 \textbf{ \textit{to} } I $}{
					\textbf{Minorization: }For a given feasible point $\overline{\gammav}^{(t-1)}$, construct a surrogate function $\breve{f}_{\text{SE}}(\overline{\gammav},\overline{\gammav}^{(t-1)})$ as given in \eqref{mm_opt_cvx}.\\
					\textbf{Maximization: } Compute $\overline{\gammav}^{(t)}$ by solving problem $\mathbf{P2}$.\\
					Repeat steps 4 and 5 until $\|\overline{\gammav}^{(t)} - \overline{\gammav}^{(t-1)} \|\leq \;\epsilon$.\\
				}
				\caption{MM Algorithm for SE optimization}
				\label{MMalgo}
			\end{algorithm}
			\begin{figure*}[t]
				\centering
				\begin{subfigure}{.24\linewidth}
					\includegraphics[width=0.86\linewidth,height=0.86\linewidth]{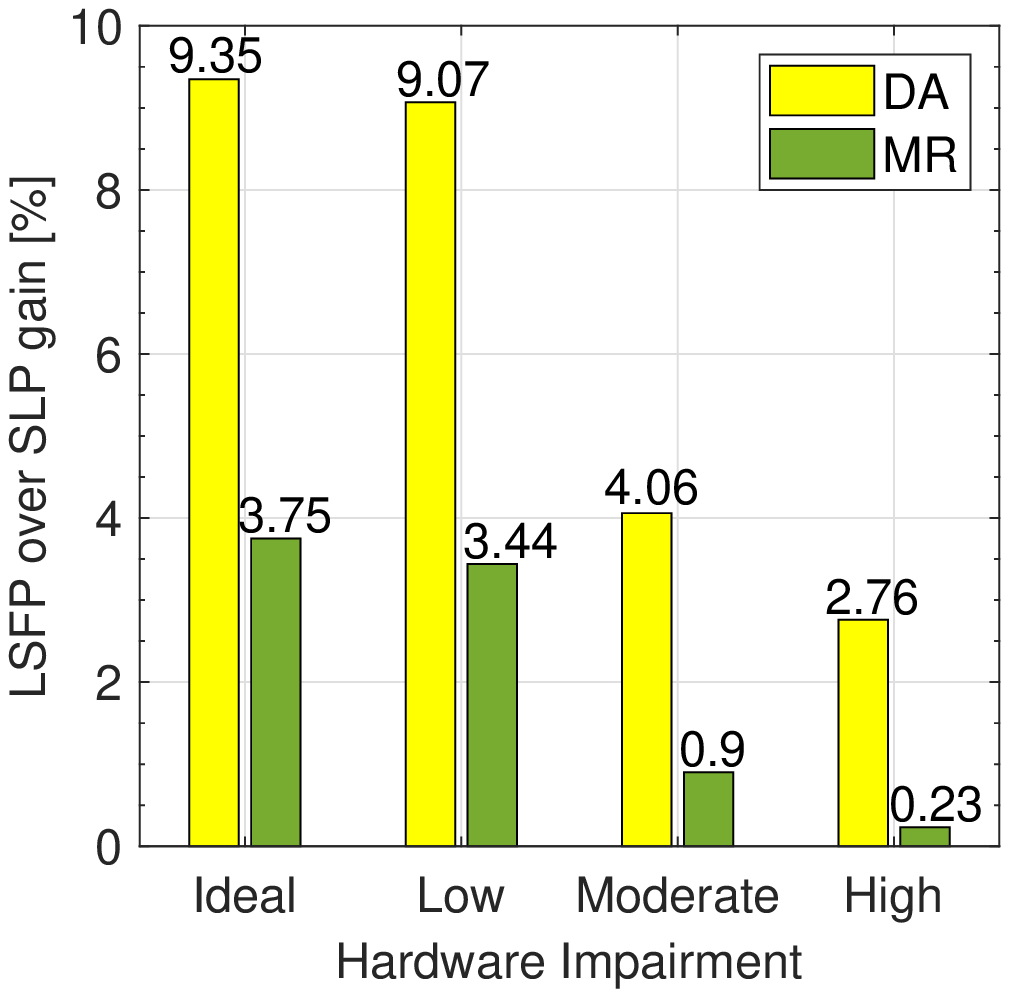}\vspace{-5pt}
					\caption{\small }\vspace{-5pt}
					\label{fig_percent_SE_LSFP_vs_SLP}
				\end{subfigure}
				\begin{subfigure}{.24\linewidth}
					\includegraphics[width=0.86\linewidth,height=0.86\linewidth]{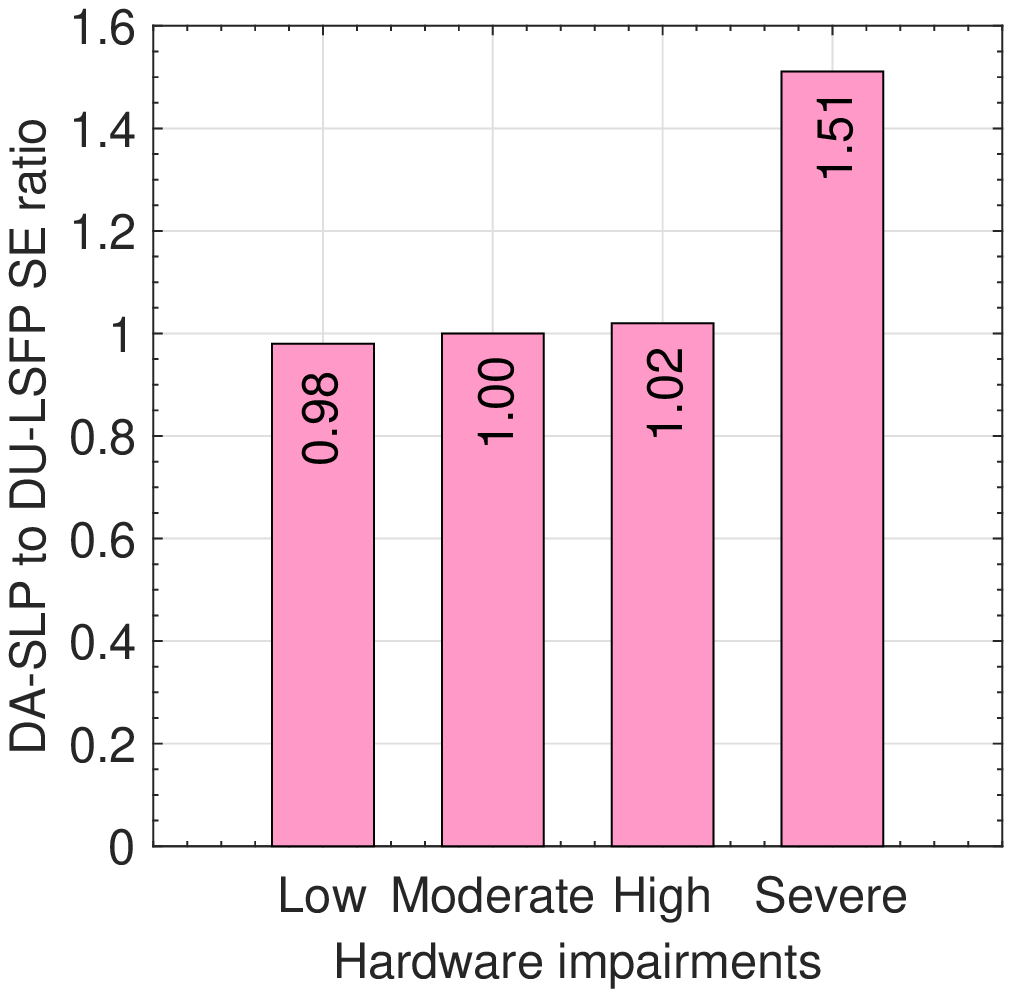}\vspace{-5pt}
					\caption{\small }\vspace{-5pt}
					\label{fig_ratio_DU_LSFP_DA_SLP}
				\end{subfigure}
				\begin{subfigure}{.24\linewidth}
					\includegraphics[width=0.86\linewidth,height=0.86\linewidth]{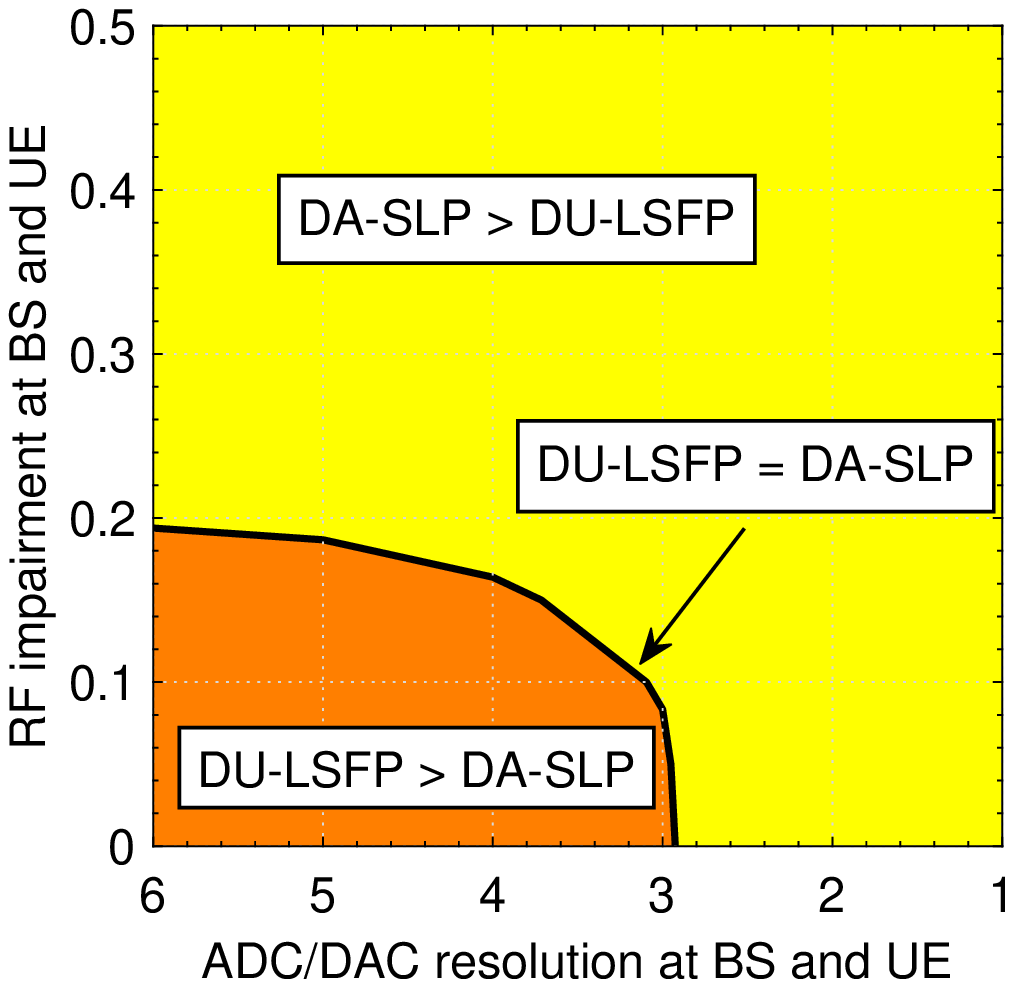}\vspace{-5pt}
					\caption{\small }\vspace{-5pt}
					\label{fig_contour_plot}
				\end{subfigure}
				\begin{subfigure}{.24\linewidth}
					\includegraphics[width=0.86\linewidth,height=0.86\linewidth]{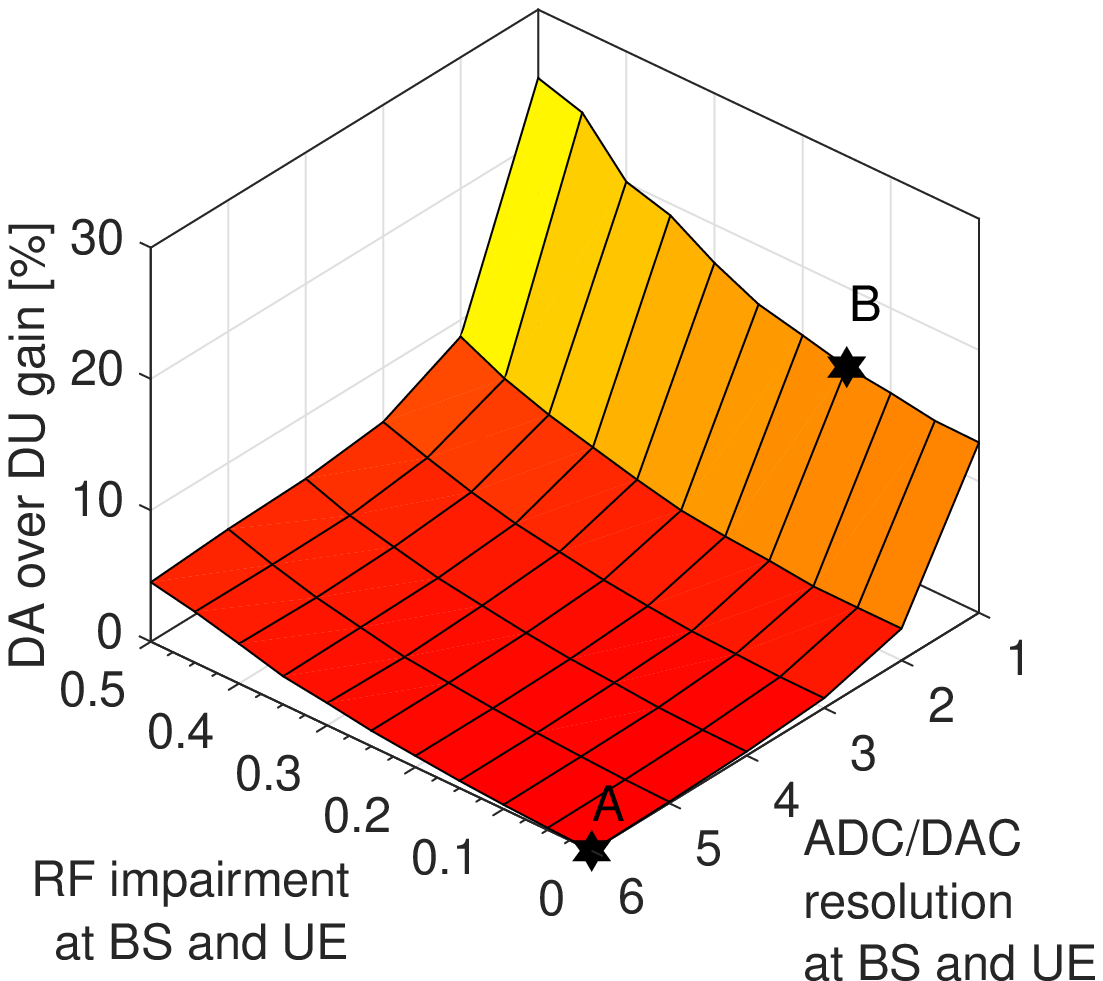}\vspace{-5pt}
					\caption{\small }\vspace{-5pt}
					\label{fig_DA_SMMSEvsDU_SMMSE}
				\end{subfigure}
				\vspace{-5pt}
				\caption{\small a) LSFP over SLP percentage SE gain and b) DA-SLP over DU-LSFP SE ratio; for different hardware impairments. c) $\kappa - b$ region depicting the regions where DA-SLP dominates over DU-LSFP and vice-versa; and d) Percentage SE gain in DA-LSFP over DU-LSFP;\\[-25pt]}
				\label{}
			\end{figure*}\vspace{-7pt}
			\section{Simulations results} \label{sec:sim_results}\vspace{-5pt}
			We now numerically analyze the performance of LSFP design, by considering a four-cell mMIMO network which is deployed in a geographical area of $1\text{Km}\times1\text{Km}$. We assume that each BS is located at the cell center, and that each UE is randomly deployed at a distance greater than $35$m from the BS.  We consider a communication bandwidth of $\mathcal{B} = 20$~MHz and a receiver noise floor of $\sigma^2 = -96$~dBm. The large-scale fading coefficients and the Rician $K$ factors are modelled as in \cite[Eq. (68)]{Ozgocan_2019}. The BS is equipped with a uniform linear array (ULA) with a half-wavelength spacing. The correlation matrices $\{\Sigmamat_{lk}^{j}\}_{\forall l,k,j}$ are modelled using Gaussian local scattering model with angular standard deviation~of~$30^\circ$~\cite{Ozgocan_2019}. The BSs use dynamic-DAC architecture with $1,2,4$ and $6$ bit resolution in equal proportion while transmitting data, and the UE ADC/DAC has $b=4$ bit resolution. The corresponding distortion factors are chosen from Table~1 in~\cite{zhang_2017}. We fix i) $K = 5$ UEs~per cell; ii) $M = 100$ BS antennas; and iii) pilot power~$\tilde{p}_{lk}=23$~dBm. The optimal LSFP coefficients~are obtained using Algorithm~\ref{MMalgo}.~We present extensive numerical results for the following designs:
			\begin{itemize}[leftmargin = *]
				\item \textit{\underline{Single-layer precoding (SLP) scheme:}} BS performs MR, DU-MMSE and DA-MMSE precoding. They are denoted as MR-SLP, DU-SLP, DA-SLP, respectively. 
				The SE for SLP scheme can be obtained by setting the LSFP coefficients as $\gamma_{lk}^{r} = 0$ for $r \neq l$ and $\gamma_{lk}^{l} = \sqrt{p_{lk}}$ in~\eqref{SE_bar}.  
				\item \textit{\underline{LSFP scheme}}: CNC peforms LSFP and the BS performs MR, DU-MMSE and DA-MMSE precoding. They are denoted as MR-LSFP, DU-LSFP, DA-LSFP, respectively.
			\end{itemize}
			\subsubsection{\textbf{Comparison of LSFP and SLP}} 
			We first investigate in Fig. \ref{fig_percent_SE_LSFP_vs_SLP}, the percentage SE gains provided by the LSFP over SLP. We perform this study when the BS employs MR and DA-MMSE precoders. We consider the  following configurations i) ideal hardware:- $\kappa_{rb}= $  $\kappa_{tb}=\kappa_{bs}= 0$, $\kappa_{ru}=$ $\kappa_{tu}= \kappa_{ue} = 0$ and  $b = \infty$ at both BS and UE; ii) low hardware impairments:- $\kappa_{bs} = 0.01$, $\kappa_{ue} =0.01$, $b = 5$ bits at UE, and  $b = (3,4,5,6)$ bits at BS in equal proportions; iii) moderate hardware impairments:- $\kappa_{bs}=0.1$, $\kappa_{ue}=0.05$, $b = 4$ bits at UE and  $b = (2,3,4,6)$ bits at BS in equal proportions; and iv) high hardware impairments:- $\kappa_{bs}=0.175$, $\kappa_{ue}= 0.1$, $b = 3$ bits at UE and  $b = (1,2,3,4)$ bits at BS in equal proportions. We crucially observe from Fig.~\ref{fig_percent_SE_LSFP_vs_SLP} that for ideal hardware, the SE gain provided by LSFP over SLP for i) DA-MMSE precoder is $9.35 \%$; and ii) MR precoder is $3.75 \%$. These numbers show that with ideal hardware, LSFP provides reasonable gain for the DA-MMSE precoder, and only moderate gain for the MR precoder. For high hardware impairments, LSFP provides $2.76\%$ gain for the DA-MMSE precoder, and $0.23\%$ gain for the MR precoder. The LFSP gain is now considerably reduced for both precoders. This  is because the interference due to hardware impairments dominate the interference due to pilot contamination. We, however, crucially note that the LSFP is still able to provide marginal gain for the DA-MMSE precoder, but only a cosmetic gain for the MR precoder. The better performance of LSFP-DA-MMSE combination is because LSFP coefficients are designed using long-term channel statistics. They are unable to effectively mitigate the interference due to hardware impairments.  The proposed DA-MMSE precoder can mitigate that, but MR cannot. With the reduced hardware impairment interference, the LSFP is able to perform better.  \textit{This study crucially informs a system designer that for practical hardware-impaired mMIMO systems, it is better to combine LSFP with DA-MMSE precoding rather with MR precoding.  The cosmetic gain provided by LSFP with MR precoding is too little to justify the LSFP complexity.}
			\subsubsection{\textbf{Comparison of DA-SLP versus DU-LSFP}}
			We now compare in Fig.~\ref{fig_ratio_DU_LSFP_DA_SLP} the SE ratio of DA-MMSE when it is used with SLP, and when DU-MMSE is used with LSFP. \textit{This study will help in investigating the use cases for LSFP.} For this study, we consider following hardware impairments i) low: $b= 5$ bits, $\kappa=0.05 $; ii) moderate: $b= 3$ bits, $\kappa=0.1$; iii) high: $b= 2$ bits, $\kappa=0.1$; and iv) severe: $(\kappa,b)=(0.15,1\text{bit})$. We see that for low impairment, the SE ratio is slightly lesser than unity. For this use case,  the DU-MMSE-LSFP combination has a marginally higher SE than DA-MMSE-SLP combination. The LSFP usage is, therefore, suggested in  this case.  For low-to-high impairment values, the SE ratio is close to unity. For severe hardware impairments, the SE ratio is $1.51$. The increased gain of DA-MMSE-SLP combination for severe hardware impairments is~due~to the ability of DA-MMSE precoder to mitigate hardware impairments, which the DU-MMSE precoder is unable to. 
			The LSFP gains for high hardware impairments with DU-MMSE precoder  are relatively lower than the DA-MMSE precoder.  A system designer should, therefore,  prefer DA-SLP for high hardware impairments, when compared with DU-LSFP precoding.
			
			
			We present in Fig. \ref{fig_contour_plot} the regions where DA-SLP design outperforms DU-LSFP design and vice-versa, for different hardware impairment values $\kappa$, and bit resolution $b$. In the orange region, the DU-LSFP outperform DA-SLP, while in the yellow region, it is the other way round.
			\emph{This study provides critical insights to the system designer regarding the use cases of DA-SLP and DU-LSFP  in  practical mMIMO systems.}
			\subsubsection{\textbf{Comparison of DA-MMSE and DU-MMSE precoders}}
			We next plot in {Fig. \ref{fig_DA_SMMSEvsDU_SMMSE}} the SE gain provided by the DA-MMSE precoder over the DU-MMSE precoder, when both of them are used with LSFP. This will help us in motivating the use of DA-MMSE precoder. For this study, we assume equal RF impairments and equal bit resolution at the BS and UE i.e., $\kappa_{bs} = \kappa_{ue} = \kappa$ and $b_{BS}$ = $b_{UE}$ = $b$. These assumptions are made only for the sake of brevity, but the observations remain same for different impairments/resolutions.  We first observe that the SE of DA-MMSE and DU-MMSE precoders match for zero RF impairments and a reasonably high ADC/DAC resolution i.e., $(\kappa,b)=(0, 6)$ (i.e., point A in Fig.~\ref{fig_DA_SMMSEvsDU_SMMSE}). This shows that DA-MMSE reduces to DU-MMSE precoder for close-to-ideal hardware, which validates its correctness. The DA-MMSE precoder gain, however, increases with $\kappa$. \textit{For $\kappa = 0.15$ and $b=1$ bit (i.e., point B in Fig.~\ref{fig_DA_SMMSEvsDU_SMMSE}), the DA-MMSE precoder has $20\%$ gain over the DU-MMSE one, which strongly motivates its use.}
			\vspace{-10pt}
			\section{Conclusion}\vspace{-7pt}
			We considered a multi-cell spatially-correlated Rician-faded mMIMO system with two-layer LSFP, dynamic-resolution ADC/DAC architecture and RF impairments at the BS and UEs. We practically modeled the Rician fading channel by including a random phase-shift in its LoS component. We derived a closed-form SE expression and proposed an algorithm to optimize the LSFP vectors to maximize the non-convex  sum-SE metric. 
			We showed that the proposed DA-MMSE precoder with single-layer processing outperforms the DU-MMSE with LSFP. However, if we need even higher gains, then LSFP with DA-MMSE is the choice.
			\appendices\vspace{-5pt}
			\section{}\vspace{-7pt} \label{app:chan_est}
			{We first tabulate in Table~\ref{tab:estimation_phase_distributions}, the mean and covariance matrices of hardware~impairments during the channel estimation phase.}
			\begin{table*}[htbp]
				\footnotesize
				\begin{tabular}{|c|c|c|c|c|c|}
					\hline
					Noise & Mean & Covariance & Noise & Mean & Covariance \\
					\hline
					\!\!$\hat{n}_{DAC_{l'k'}} = \bar{\nv}_{DAC_{l'k'}}^T \boldsymbol{\phi}_k^{*}$\!\! &\!\! 0 \!\!&\!\! $ \tau_p(1-\alpha_{d_{l'k'}}) \alpha_{d_{l'k'}} \tilde{p}_{l'k'}$\!\! &\!\!$\hat{\eta}_{tu_{l'k'}}=\bar{\boldsymbol{\eta}}_{tu_{l'k'}}^T \boldsymbol{\phi}_k^{*}$\!\! & \!\!$0$\!\! &\!\! $ \tau_p\kappa_{tu}^2 \alpha_{d_{l'k'}} \tilde{p}_{l'k'}$\!\! \\
					\hline
					\!\!$\hat{\boldsymbol{\eta}}_{rb}^{j}  =\overline{\boldsymbol{\eta}}_{rb}^{j} \boldsymbol{\phi}_k^{\ast}$\!\! & \!\!$\mathbf{0}_{M\times 1}$\!\! & \!\!$\tau_p \kappa_{bs}^2 \overline{\mathbf{W}}^{j}$\!\! & \!\!$\bar{\nv}_{ADC}^j=\overline{\mathbf{N}}_{q}^{j} \boldsymbol{\phi}_k^{*}$\!\! &\!\!$\mathbf{0}_{M \times 1}$\!\! &\!\! $\tau_p \mathbf{B}^{j} ((1+\kappa_{bs}^2) \overline{\mathbf{W}}^{j} + \sigma^2\mathbf{I}_M)$\!\!\\
					\hline
					\multicolumn{6}{|c|}{Here $\overline{\mathbf{W}}^{j} = \sum\limits_{l'=1}^{L} \sum\limits_{k'=1}^{K} (1 + \kappa_{tu}^2) \alpha_{d_{l'k'}} \tilde{p}_{l'k'} \text{diag}\{\mathbf{h}_{l'k'}^{j}\mathbf{h}_{l'k'}^{j^H}\} + \sum\limits_{l'=1}^{L} \sum\limits_{m \neq l'}^{L} \sum\limits_{k'=1}^{K} \alpha_{d_{l'k'}} \alpha_{d_{mk'}} \sqrt{\tilde{p}_{l'k'}\tilde{p}_{mk'}} \text{diag}\{\mathbf{h}_{l'k'}^{j}\mathbf{h}_{mk'}^{j^H}\}  $.}\\
					\hline
				\end{tabular}
				\caption{Statistical parameters of different distortion/quantization noise .\vspace{-7pt}} \label{tab:estimation_phase_distributions}
			\end{table*}
			We recall that the received pilot signal is given~as\vspace{2pt}
			\begin{align} \label{pilot_rx_sig}
			\mathbf{y}_{jk}	&= \underbrace{\sum_{l=1}^{L}\alpha_{d_{lk}}\sqrt{\tilde{p}_{lk}}\tau_{p}\A_a^j\h_{lk}^j\!}_{\mathbf{f}_{jk}^1}+\underbrace{\!\sum_{l=1}^L\sum_{k=1}^K\A_a^j\h_{lk}^j(\hat{n}_{DAC_{lk}}\!+\!\hat{\eta}_{tu_{lk}})\!}_{\mathbf{f}_{jk}^2}\nonumber\\[-5pt]
				&+\underbrace{\!\A_a^j\hat{\boldsymbol{\eta}}_{rb}^j\!+\!\A_a^j\overline{\mathbf{z}}^j\!+\!\overline{\mathbf{n}}_{ADC}^j}_{\mathbf{f}_{jk}^3}. \tag{A1}
			\end{align}
			The phase unaware LMMSE channel estimate is given as~[4]:
			\begin{align}\tag{A2}\label{eq_CH_estimate9}
				&\hat{\mathbf{h}}_{lk}^{j} = \mathbf{C}_{\mathbf{h}_{lk}^{j}\mathbf{y}_{jk}}\mathbf{C}_{\mathbf{y}_{jk}\mathbf{y}_{jk}}^{-1} \mathbf{y}_{jk}
			\end{align}
		where $\mathbf{C}_{\mathbf{h}_{lk}^{j} \mathbf{y}_{jk}} = \E\{\mathbf{y}_{jk}\mathbf{h}_{lk}^{j^H}\}$ and $\mathbf{C}_{\mathbf{y}_{jk}\mathbf{y}_{jk}} = \E\{\mathbf{y}_{jk}\mathbf{y}_{jk}^H\}$.

			{The matrices $\Cmat_{\hv_{lk}^{j}\yv_{jk}}$ and $\Cmat_{\yv_{jk}\yv_{jk}}$ can be simplified by substituting $\yv_{jk}$ from \eqref{pilot_rx_sig}, and by using the statistics of distortion/quantization noises from Table~\ref{tab:estimation_phase_distributions}. Their final expressions are given as:}
			\begin{align}
				&\mathbf{C}_{\mathbf{h}_{lk}^{j} \mathbf{y}_{jk}} = \alpha_{d_{lk}}\tau_p\sqrt{\tilde{p}}_{lk}\A^j_a\overline{\R}_{lk}^j \text{ with } \overline{\Rmat}_{lk}^{j} = \big(\Rmat_{lk}^{j} + \bar{\hv}_{lk}^j\bar{\hv}_{lk}^{j^H}\big), \nonumber\\[-2pt] 
				&\mathbf{C}_{\mathbf{y}_{jk}\mathbf{y}_{jk}}=\sum_{l'=1}^L \alpha_{d_{lk}}^2\tau_p^2\tilde{p}_{lk}\mathbf{A}^{j}_a\overline{\mathbf{R}}_{l'k}^j\mathbf{A}^{j^H}_a +\tau_p\kappa_{bs}^2\A^j_a{\mathbf{J}}^j\A^{j^H}_a\nonumber\\
				&\qquad+\sum_{l'=1}^L\sum_{k'=1}^K \tau_p\alpha_{d_{l'k'}}(1-\alpha_{d_{l'k'}}+\kappa_{tu}^2)\tilde{p}_{l'k'}\mathbf{A}^j_a\overline{\mathbf{R}}_{l'k'}^j\mathbf{A}^{j^H}_a \nonumber\\[-2pt]
				&\qquad +\sigma^2\tau_p \A^{j^2}_a+\tau_p\mathbf{B}^j_a\left(\left(1+\kappa_{bs}^2\right){\mathbf{J}}^j+\sigma^2\mathbf{I}_M)\right),\tag{A4}\label{C_yy}
			\end{align} 
		$\text{ with } \mathbf{J}^j = \E\big\{\overline{\Wmat}^j \big\}$.
			Substituting \eqref{C_yy} in \eqref{eq_CH_estimate9}, we obtain the final expression for $\hat{\hv}_{lk}^{j}$ as in~(5) of the paper.\vspace{-7pt}
			\section{}\label{app:DAMMSE}\vspace{-5pt}
			We similar to [3], [4], first design the uplink distortion-aware combiner and later use it to obtain the downlink precoder. We now use \eqref{pilot_rx_sig} to write the uplink received signal as in \eqref{LSFD_brief} shown at the top of next page
			\begin{figure*}[t]\vspace{-7pt}
				{\small\begin{align}\label{LSFD_brief} 
						\!\!\!\!y_{lk} &= \underbrace{\vv^{H}_{lk}\!\mathbf{A}_{a}^{l}\mathbf{h}_{lk}^{l}\alpha_{d_{lk}}\sqrt{\tilde{p}_{lk}}s_{lk}}_{\text{desired signal}}+\underbrace{\!\sum_{\substack{ l'\neq l}}^{L}\vv^{H}_{lk}\mathbf{A}_{a}^l\mathbf{h}_{l'k}^{l}\alpha_{d_{l'k}}\sqrt{\tilde{p}_{l'k}}s_{l'k}}_{\text{pilot contamination}} +\underbrace{\!\sum_{l'=1}^{L}\sum_{\substack{k'\neq k}}^K \vv^{H}_{lk}\mathbf{A}_{a}^l\mathbf{h}_{l'k'}^{l}\alpha_{d_{l'k'}}\sqrt{\tilde{p}_{l'k'}}s_{l'k'}}_{\text{non-pilot co-channel interference}} \nonumber\\
						&\hspace{10pt}+\underbrace{\sum_{l'=1}^{L}\sum_{\substack{k'=1}}^K \vv^{H}_{lk}\mathbf{A}_{a}^l\mathbf{h}_{l'k'}^{l}n_{\!DAC_{l'k'}}}_{\text{DAC impairments}} + \underbrace{\sum_{l'=1}^{L}\!\sum_{\substack{k'=1}}^K \vv^{H}_{lk}\mathbf{A}_{a}^{l}\mathbf{h}_{l'k'}^{l}\eta_{tu_{l'k'}}}_{\text{UE transmit RF impairments}}+\underbrace{\vv^{H}_{jk}\mathbf{A}_{a}^{l}\boldsymbol{\eta}_{bs}^{l}}_{\substack{\text{BS receive} \\ \text{RF impairments}}}+\underbrace{\vv^{H}_{lk}\mathbf{n}_{q}^l}_{\substack{\text{ADC}\\ \text{impairments}}} +\underbrace{\vv^{H}_{lk}\mathbf{A}_{a}^{\!l}\mathbf{z}^{l}}_{\text{noise}} \tag{B1}\!.
					\end{align}
			\begin{align}\label{uplink SINR} 
				&\text{SINR}_{lk}^{\text{UL}} =  \frac{\big|\alpha_{d_{lk}}\sqrt{p_{lk}}\vv_{lk}^{H}\Amat_{a}^{l}\hat{\hv}_{lk}^l\big|^2}
				{\begin{Bmatrix}						\sum\limits_{l'=1}^{L}\sum\limits_{k'=1}^{K} \!\tilde{\alpha}_{d_{l'k'}} \tilde{p}_{l'k'}  \vv_{lk}^{H}\Amat_{a}^{l}\big(\hat\hv_{l'k'}^{l}\hat{\hv}_{l'k'}^{l^H}\!+\!\Cmat_{l'k'}^{l}\big)\Amat_{a}^{l^H}\vv_{lk} 
						\!-\! \alpha_{d_{lk}}^2 \tilde{p}_{lk} \big|\vv_{lk}^{H}\Amat_a^{l}\hat{\hv}_{lk}^{l}\big|^2\!\!\!  \\[-3pt]
						+ \kappa_{rb}^2 \vv_{lk}^{H}\Amat_a^{l}{\Wmat}^{l}\Amat_a^{H}\vv_{lk}\!\!+ \vv_{lk}^{H} \Bmat_{a}^{l}{\Smat}^{l}\vv_{lk} + \sigma^2 \left|\vv_{lk}^{H}\Amat_a^{l}\right|^2\!
					\end{Bmatrix}
				}\eqa \frac{\big|\vv_{lk}^{l^H}\tilde{\av}_{lk}\big|^2}{\vv_{lk}^H\widetilde{\Bmat}_{lk}\vv_{lk}}. \tag{B2}
				\end{align}}
				\hrule
				\vspace{-13pt}
				\setcounter{equation}{12}
			\end{figure*} 

			{We obtain this expression by i) considering the $m$th column of the receive pilot signal $\yv_{m}^{j} \triangleq \Ymat^{j}(:,m)$ in \eqref{pilot_rx_sig}; ii) replacing pilot symbol $\phiv_{k_m}$ with data symbol $s_{lk}$ and; iii) combining the resultant signal with $\vv_{lk}$} as $y_{lk} = \vv_{lk}^{H}\yv_{m}^{j}$. The SINR in uplink transmission using \eqref{LSFD_brief} is written~as in \eqref{uplink SINR} shown at the top of the page.
		Here $\tilde{\av}_{lk} = \alpha_{d_{lk}} \sqrt{\tilde{p}_{lk}} \Amat_a^{l}\hat\hv_{lk}^{l}$ and
			$\widetilde{\Bmat}_{lk} = \sum_{k'=1}^{K} \tilde{\alpha}_{d_{lk'}}  \tilde{p}_{lk'} \Amat_a^{l}\big(\Cmat_{lk'}^{l} + \hat{\hv}_{lk'}^{l}\hat{\hv}_{lk'}^{l^H}\big)\Amat_a^{l^H} + \sum_{l' \neq l}^{L}\sum_{k'=1}^{K} \tilde{\alpha}_{d_{l'k'}}  \tilde{p}_{l'k'} \Amat_a^{l}\overline{\Rmat}_{l'k'}^{l}$ $\Amat_{a}^{l^H} - \alpha_{d_{lk}}^2 \tilde{p}_{lk} \Amat_a^{l}\hat\hv_{lk}^{l}\hat\hv_{lk}^{l^H}\Amat_a^{l^H} + \kappa_{rb}^2 \Amat_a^{l}\widehat{\Wmat}^{l}\Amat_a^{l^H} + \Bmat_{a}^{l}\widehat{\Smat}^{l}$ $ + \sigma^2 \Amat_a^{l}\Amat_a^{l^H}$, with $\tilde{\alpha}_{d_{lk}} = \alpha_{d_{lk}}(1+\kappa_{tu}^2)$.
			Here $\widehat{\Wmat}^{l} = \sum_{k'=1}^{K} \tilde{\alpha}_{d_{lk'}}\tilde{p}_{lk'} \text{diag}(\hat\hv_{lk'}^{j}\hat\hv_{lk'}^{j^H} + \Cmat_{lk'}^{j}) + \sum_{l'\neq l}^{L}\sum_{k'=1}^{K}\tilde{\alpha}_{d_{l'k'}}\tilde{p}_{l'k'} \text{diag}(\overline{\Rmat}_{l'k'}^{j})$ and $\widehat{\Smat}^{l} = (1 + \kappa_{rb}^2)\widehat{\Wmat}^{l} + \sigma^2\Imat_M$. 
			The matrix $\Cmat_{lk}^{l}$ is the error covariance matrix, given as $\Cmat_{lk}^{l} = \overline{\Rmat}_{lk}^{l}-\tau_p \tilde{p}_{lk}\alpha_{d_{lk}}^2 \Amat_a^{r}\overline{\Rmat}_{lk}^{l}\Psimat_{lk}^{-1}\overline{\Rmat}_{lk}^{l}\Amat_a^{r^H}$. 
			
			Equality $(a)$ in \eqref{uplink SINR} is valid because each BS uses only the channel estimates of UEs in its cell and long-term channel statistics of UEs in other cells. We, therefore, replace the terms $\big(\hat{\hv}_{l'k'}^{l^H}+ \Cmat_{l'k'}^{l}\big)$ for $l' \neq l$, with its statistics $\overline{\Rmat}_{l'k'}^{l}$.
			We note that $\text{SINR}^{\text{UL}}_{lk}$ is in generalized Rayleigh quotient form. The optimal combiner $\vv_{lk}$ is, therefore,  given as: 
			\begin{align}
			\!\!\!\!\vv_{lk}^{\text{da}} \!&=\! \widetilde{\Bmat}^{-1}_{lk}\tilde{\av}_{lk} \!\stackrel{(a)}{=}\! \left(1 \!+\! \tilde{\av}_{lk}^H\widetilde{\Bmat}_{lk}^{-1}\tilde{\av}_{lk}\right)\left(\widetilde{\Bmat}_{lk} \!+\!\tilde{\av}_{lk}\tilde{\av}_{lk}^H \right)^{-1}\!\!\tilde{\av}_{lk}.\tag{B3}
			\end{align}
			Equality $(a)$ is due to the matrix identity~ $(\Amat + \xv\xv^H)^{-1} \xv = \frac{1}{(1+\xv^H\Amat^{-1}\xv)}\Amat^{-1}\xv$. 
			The optimal distortion-aware precoder is obtained using $\vv_{lk}^{\text{DA-MMSE}}$, and by writing it in compact form~as 
			\begin{align} 
			\wv_{lk}^{l} \!&=\! \omega_{lk} \vv_{lk}^{\text{da}},\; \text{ where } \Vmat_{l}^{\text{da}} \!=\! [\vv_{l1}^{\text{da}},\cdots,\vv_{lK}^{\text{da}}] \!=\! \widehat{\Vmat}_l^{-1} \widehat{\Hmat}_l \overline{\Pmat}_l \nonumber
			\end{align}
			Here $\widehat{\Hmat}_l \!=\! [\hat{\hv}_{l1}^{l},\cdots,\hat{\hv}_{lK}^{l}]$, $\overline{\Pmat}_l = \text{diag}(\tilde{p}_{l1},\cdots,\tilde{p}_{lK})$~and
			$\widehat{\Vmat}_{l} = \sum\limits_{k'=1}^{K} \tilde{\alpha}_{d_{lk'}} \tilde{p}_{lk'} \big(\hat{\hv}_{lk'}^{l}\hat{\hv}_{lk'}^{l^H} + \Cmat_{lk'}^{l}\big)\Amat_a^{l^H} + \sum\limits_{l' \neq l}^{L}\sum\limits_{k'=1}^{K}\tilde{\alpha}_{d_{l'k'}}\tilde{p}_{l'k'}$ $  \overline{\Rmat}_{l'k'}^{l}\Amat_{a}^{l^H}$ $+ \kappa_{rb}^2 \widehat{\Wmat}^{l}\Amat_a^{l^H}+  (\Imat_M - \Amat_a^{l})\widehat{\Smat}^{l} + \sigma^2\Amat_a^{l^H}$.
			\vspace{-5pt}
			\section{}\label{app:MM_cond}
			\vspace{-5pt}
			We first rewrite the fractional term ${A_{lk}(\xv)}/{B_{lk}(\xv)}$ as $\frac{N_{lk}(\xv,y_k)^2}{D_{lk}(\xv,y_k)} \triangleq \frac{y_{lk}^2\sqrt{A_{lk}(\xv)}^2}{ y_{lk}^2B_{lk}(\xv)}$. Using the arithmetic mean-harmonic mean inequality over the functions $N_{lk}(\xv,y_{lk})$, $D_{lk}(\xv,y_{lk})$, we have $\left[N_{lk}(\xv,y_{lk}) + D_{lk}(\xv,y_{lk})\right]/{2} \geq {2}/{\left[\frac{1}{N_{lk}(\xv,y_{lk})}+ \frac{1}{D_{lk}(\xv,y_{lk})}\right]}$ ${\stackrel{(a)}{\Rightarrow} \frac{N_{lk}(\xv,y_{lk})^2}{D_{lk}(\xv,y_{lk})} \geq 2N_{lk}(\xv,y_{lk}) - D_{lk}(\xv,y_{lk})}$ ${\stackrel{(b)}{\Rightarrow} \frac{A_{lk}(\xv)}{B_{lk}(\xv)} \geq 2y_{lk}\sqrt{A_{lk}(\xv)} - y_{lk}^2B_{lk}(\xv)\triangleq h_{lk}(\xv,y_{lk})}$.
			\newline
			Implication $(a)$ is obtained by rearranging the terms, and implication $(b)$ is obtained by substituting $N_{lk}(\xv,y_{lk})  = y_{lk}\sqrt{A_{lk}(\xv)}$ and $D_{lk}(\xv,y_{lk})  = y_{lk}^2\sqrt{B_{lk}(\xv)}$. The scalar $y_{lk}$ is designed such that it satisfies the conditions in (17). As per the first condition in (17), $h_{lk}(\xv,y_{lk})\big|_{\xv = \xv_t} = \frac{A_{lk}(\xv_t)}{B_{lk}(\xv_t)}\Rightarrow 2y_{lk}\sqrt{A_{lk}(\xv_t)} - y_{lk}^2B_{lk}(\xv_t) = \frac{A_{lk}(\xv_t)}{B_{lk}(\xv_t)}$.\newline
		On arranging these terms, we obtain $y_{lk} = \frac{\sqrt{A_{lk}(\xv_t)}}{B_{lk}(\xv_t)}$.
			\vspace{-5pt}
			\bibliographystyle{IEEEtran}
			\bibliography{IEEEabrv,lsfp_ref}

\begin{thebibliography}{10}
\providecommand{\url}[1]{#1}
\csname url@samestyle\endcsname
\providecommand{\newblock}{\relax}
\providecommand{\bibinfo}[2]{#2}
\providecommand{\BIBentrySTDinterwordspacing}{\spaceskip=0pt\relax}
\providecommand{\BIBentryALTinterwordstretchfactor}{4}
\providecommand{\BIBentryALTinterwordspacing}{\spaceskip=\fontdimen2\font plus
\BIBentryALTinterwordstretchfactor\fontdimen3\font minus
  \fontdimen4\font\relax}
\providecommand{\BIBforeignlanguage}[2]{{%
\expandafter\ifx\csname l@#1\endcsname\relax
\typeout{** WARNING: IEEEtran.bst: No hyphenation pattern has been}%
\typeout{** loaded for the language `#1'. Using the pattern for}%
\typeout{** the default language instead.}%
\else
\language=\csname l@#1\endcsname
\fi
#2}}
\providecommand{\BIBdecl}{\relax}
\BIBdecl

\bibitem{Ozgocan_2019}
{\"O}.~{\"O}zdogan, E.~Bj{\"o}rnson, and E.~G. Larsson, ``Massive {MIMO} with
  spatially correlated rician fading channels,'' \emph{{IEEE} Trans. Commun.},
  vol.~67, no.~5, pp. 3234--3250, 2019.

\bibitem{Ashikhmin_2018}
A.~Ashikhmin, L.~Li, and T.~L. Marzetta, ``Interference reduction in multi-cell
  massive {MIMO} systems with large-scale fading precoding,'' \emph{{IEEE}
  Trans. Inf. Theory}, vol.~64, no.~9, pp. 6340--6361, 2018.

\bibitem{demir_large2020}
{\"O}.~T. Demir and E.~Bj{\"o}rnson, ``Large-scale fading precoding for
  spatially correlated rician fading with phase shifts,'' \emph{arXiv preprint
  arXiv:2006.14267}, 2020.

\bibitem{Ozdogan_2019}
O.~Ozdogan, E.~Bjornson, and J.~Zhang, ``Performance of cell-free massive
  {MIMO} with rician fading and phase shifts,'' \emph{{IEEE} Trans. Wireless
  Commun.}, vol.~18, no.~11, pp. 5299--5315, 2019.

\bibitem{Van_2019}
T.~Van~Chien, C.~Mollén, and E.~Björnson, ``Large-scale-fading decoding in
  cellular massive {MIMO} systems with spatially correlated channels,''
  \emph{{IEEE} Trans. Commun.}, vol.~67, no.~4, pp. 2746--2762, 2019.

\bibitem{bjornson_da_2018}
E.~Bj{\"o}rnson, L.~Sanguinetti, and J.~Hoydis, ``Hardware distortion
  correlation has negligible impact on {UL} massive {MIMO} spectral
  efficiency,'' \emph{{IEEE} Trans. Commun.}, vol.~67, no.~2, pp. 1085--1098,
  2018.

\bibitem{liu_2020}
T.~Liu, J.~Tong, Q.~Guo, J.~Xi, Y.~Yu, and Z.~Xiao, ``On the performance of
  massive {MIMO} systems with low-resolution {ADC}s and {MRC} receivers over
  rician fading channels,'' \emph{{IEEE} Syst. J.}, vol.~15, no.~3, pp.
  4514--4524, 2020.

\bibitem{jacobsson_2017}
S.~Jacobsson, G.~Durisi, M.~Coldrey, U.~Gustavsson, and C.~Studer, ``Throughput
  analysis of massive {MIMO} uplink with low-resolution {ADC}s,'' \emph{{IEEE}
  Trans. Wireless Commun.}, vol.~16, no.~6, pp. 4038--4051, 2017.

\bibitem{xu_2019}
L.~Xu, X.~Lu, S.~Jin, F.~Gao, and Y.~Zhu, ``On the uplink achievable rate of
  massive {MIMO} system with low-resolution {ADC} and {RF} impairments,''
  \emph{{IEEE} Commun. Lett.}, vol.~23, no.~3, pp. 502--505, 2019.

\bibitem{liang_2016}
N.~Liang and W.~Zhang, ``Mixed-{ADC} massive {MIMO},'' \emph{{IEEE} J. Sel.
  Areas Commun.}, vol.~34, no.~4, pp. 983--997, 2016.

\bibitem{pirzadeh_2018}
H.~Pirzadeh and A.~L. Swindlehurst, ``Spectral efficiency of mixed-{ADC}
  massive {MIMO},'' \emph{{IEEE} Trans. Signal Process.}, vol.~66, no.~13, pp.
  3599--3613, 2018.

\bibitem{zhang_2017}
J.~Zhang, L.~Dai, Z.~He, S.~Jin, and X.~Li, ``Performance analysis of
  mixed-{ADC} massive {MIMO} systems over rician fading channels,''
  \emph{{IEEE} J. Sel. Areas Commun.}, vol.~35, no.~6, pp. 1327--1338, 2017.

\bibitem{tugfe2019channel}
{\"O}.~T. Demir and E.~Bj{\"o}rnson, ``Channel estimation in massive {MIMO}
  under hardware non-linearities: Bayesian methods versus deep learning,''
  \emph{IEEE Open J. Commun. Soc.}, vol.~1, pp. 109--124, 2020.

\bibitem{sun_2017}
Y.~Sun, P.~Babu, and D.~P. Palomar, ``Majorization-minimization algorithms in
  signal processing, communications, and machine learning,'' \emph{{IEEE}
  Trans. Signal Process.}, vol.~65, no.~3, pp. 794--816, 2017.

\bibitem{Jeon_Mag_2021}
J.~Jeon, G.~Lee, A.~A. Ibrahim, J.~Yuan, G.~Xu, J.~Cho, E.~Onggosanusi, Y.~Kim,
  J.~Lee, and J.~C. Zhang, ``{MIMO} evolution toward 6{G}: Modular massive
  {MIMO} in low-frequency bands,'' \emph{{IEEE} Commun. Mag.}, vol.~59, no.~11,
  pp. 52--58, 2021.

\bibitem{Demir_2021}
O.~T. Demir and E.~Bjornson, ``The bussgang decomposition of nonlinear systems:
  Basic theory and {MIMO} extensions [lecture notes],'' \emph{{IEEE} Signal
  Process. Mag.}, vol.~38, no.~1, pp. 131--136, 2021.

\bibitem{Boyd_2014}
S.~Boyd and L.~Vandenberghe, \emph{Convex optimization}.\hskip 1em plus 0.5em
  minus 0.4em\relax Cambridge university press, 2004.

\end{thebibliography}
		\end{document}